\documentclass[aps, prx, twocolumn, superscriptaddress, amsmath, tightenlines, longbibliography,nofootinbib]{revtex4-2}

\newcommand{\RNum}[1]{\uppercase\expandafter{\romannumeral #1\relax}}
\usepackage{color}

\usepackage{amssymb}
\usepackage{amsmath}
\usepackage{dcolumn}
\usepackage{graphicx}
\usepackage{mathrsfs}
\usepackage{appendix}
\usepackage{graphicx}
\usepackage{booktabs}
\usepackage{colortbl}
\usepackage{braket}
\usepackage{array}
\usepackage{mleftright}

\definecolor{Dred}{RGB}{190,0,0}

\newcommand{\figpanel}[2]{Fig.~\hyperref[#1]{\ref*{#1}(#2)}}
\newcommand{\Figurepanel}[2]{Figure~\hyperref[#1]{\ref*{#1}(#2)}}
\newcommand{\figpanels}[3]{Fig.~\hyperref[#1]{\ref*{#1}(#2)-(#3)}}
\newcommand{\figpanelNoPrefix}[2]{\hyperref[#1]{\ref*{#1}(#2)}}

\newcommand{\secref}[1]{\mbox{Sec.~\ref{#1}}}

\usepackage{url}
\usepackage[colorlinks]{hyperref}
\hypersetup{%
	plainpages=true,
	breaklinks=true,       
	hypertexnames=false,  
	pageanchor=true,
	colorlinks=true,
	linkcolor={blue},
	citecolor={red},
	urlcolor={blue},
	anchorcolor={black}
}

\def \hide#1{}

\begin{document}

\title{Generating spatially separated correlated multiphoton states \\ in nonlinear waveguide quantum electrodynamics}

\author{Jia-Qi Li}
\affiliation{Shaanxi Province Key Laboratory of Quantum Information and 
Quantum Optoelectronic Devices, School of Physics, Xi'an Jiaotong University, 
Xi'an 710049, People's Republic of China}

\author{Anton Frisk Kockum}
\affiliation{Department of Microtechnology and Nanoscience, Chalmers University of Technology, 41296 Gothenburg, Sweden}

\author{Xin Wang}\email{wangxin.phy@xjtu.edu.cn}
\affiliation{Shaanxi Province Key Laboratory of Quantum Information and 
Quantum Optoelectronic Devices, School of Physics, Xi'an Jiaotong University, 
Xi'an 710049, People's Republic of China}

\date{\today}


\begin{abstract}
Strongly correlated multi-photon states are indispensable resources for advanced quantum technologies, yet their deterministic generation remains challenging due to the inherent weak nonlinearity in most optical systems. Here, we propose a scalable architecture for producing correlated few-photon entangled states via cascaded inelastic scattering in a nonlinear waveguide. When a single photon scatters off a far detuned excited two-level emitter, it coherently converts into a propagating doublon, a bound photon pair with anomalous dispersion. This doublon can subsequently scatter off a downstream excited emitter to further convert into a triplon, and so on, thereby establishing a photon-number amplification cascade $|\cdot \rangle \!\! \rightarrow \!\! |\!\!: \rangle \!\! \rightarrow \! \! |\!\!\therefore \rangle \!\! \to \!\! ...$ Central to this process is the concept of a pseudo-giant atom, which we introduce here to capture the non-local scattering potential emergent from the wave functions of bound states. By implementing this scheme using a real giant atom with multiple engineered coupling points, we achieve unidirectional and full controllable photon conversion without backscattering. The resulting output state forms a programmable superposition of spatially and temporally isolated photon-number components, automatically sorted by their distinct group velocities. This work opens a new paradigm in quantum state engineering, enabling on-demand generation of complex multi-photon resources for quantum simulation, metrology, and scalable quantum networks.	
\end{abstract}

\maketitle


\section{Introduction}

Strongly correlated multi-photon states are fundamental to modern quantum technologies, enabling quantum-enhanced metrology, scalable quantum networks, and quantum simulation of many-body physics~\cite{Rolston2002, Chang2014, Bloch2022}. However, generating such states often requires strong optical nonlinearities at the few-photon level, which remains a significant challenge due to the inherently weak photon-photon interactions~\cite{Roy2017}. Conventional approaches for generating multi-photon states in waveguide quantum electrodynamics (wQED) leverage two-level emitters (TLEs) as nonlinear impurities with infinite on-site interaction ($U\to\infty$), but the limited optical depth constrains the efficiency of the multi-photon state generation~\cite{Chang2018, Sheremet2023, Alejandro2024}. Alternative strategies employing extended emitter ensembles~\cite{Prasad2020, Poshakinskiy2021, CalajoG2022} or strong coupling~\cite{Peyronel2012, Lodahl2015, LeJeannic2022} face challenges in accuracy, control, and scalability. 

Another alternative route to generation of multi-photon states is nonlinear photonic baths. Here, the nonlinearity is directly embedded into the bath, described by a Kerr-type or Bose--Hubbard Hamiltonian. In such systems, photons directly experience pairwise interactions, leading to the formation of multi-photon bound states~\cite{Mansikkamaki2022}. In experimental platforms for quantum simulation~\cite{Schfer2020}, such Hamiltonians arise naturally, e.g., from the intrinsic anharmonicity of the transmon qubit in circuit QED~\cite{Koch2007, Gu2017, Orell2019, Blais2021, Fedorov2021, Karamlou2024} and the dipole blockade in Rydberg atom arrays~\cite{Lukin2001, Gorshkov2011, Weckesser2024}. Crucially, these multi-photon bound states,such as doublons (two-photon states) and triplons (three-photon states), inherit dispersion relations different from the linear multi-photon scattering state, behaving as quasi-particles with defined velocity and energy~\cite{Piil2007, Valiente2008, Valiente2010}. However, dynamically generating, manipulating, and spatially isolating these complex states within a single platform remains elusive.

\begin{figure*}
\centering \includegraphics[width=0.9\linewidth]{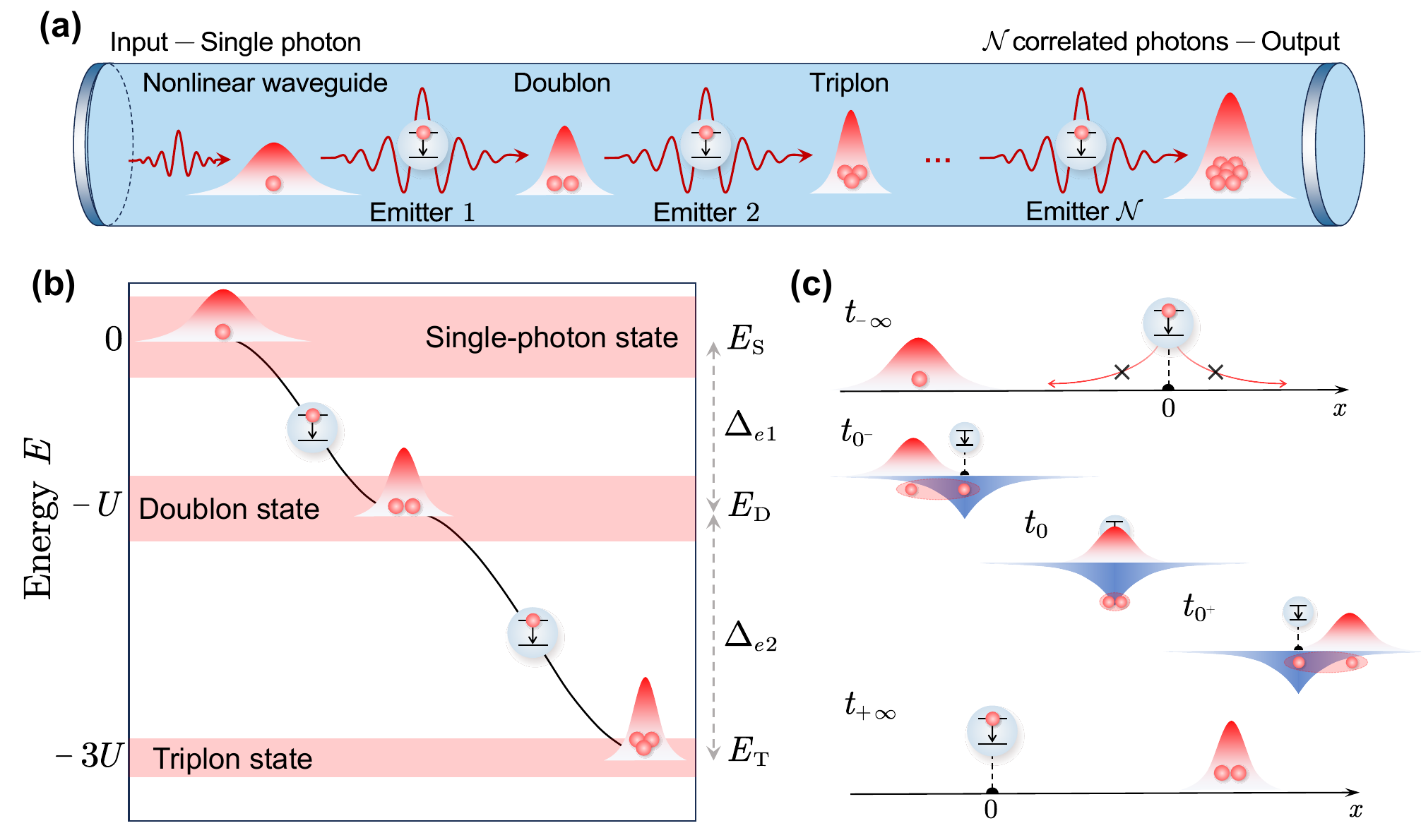}
\caption{Schematic of cascaded scattering in a nonlinear waveguide.
(a) A single photon incident from the left in the waveguide first scatters off a detuned excited emitter, generating a pair of correlated photons---a doublon. The doublon subsequently scatters off a second excited emitter, producing three correlated photons---a triplon. This cascaded scattering process can, in principle, be extended to generate $\mathcal{N}$ correlated photons.
(b) Energy-level structure for single- ($E_{\mathrm{S}}$), two- ($E_{\mathrm{D}}$), and three-photon ($E_{\mathrm{T}}$) wave packets and the emitters with transition frequencies detuned by $\Delta_{e1}$ and $\Delta_{e2}$ from the waveguide cavity frequency $\omega_c$. The bandwidth of the single-photon band is 4J; the doublon and triplon bands are more narrow than that, and are not shown exactly to scale in this illustration.
(c) Temporal evolution of the scattering process. The shading depth at times $\{t_{0^-}, t_0, t_{0^+}\}$ represents the instantaneous effective scattering strength.}
	\label{fig1}
\end{figure*}

Here, we introduce a cascaded inelastic scattering process within a nonlinear waveguide that overcomes these limitations. Our approach exploits far detuned excited TLEs as non-resonant mediators, enabling a controlled sequential upconversion of photon-number correlations [see \figpanel{fig1}{a}]. Crucially, an incident single photon ($S$) scatters off the first excited TLE to deterministically generate a propagating doublon ($D$), a bound photon pair with unique dispersion. This doublon state then acts as the input for a second scattering event off a subsequent excited TLE, coherently upconverting it into a triplon ($T$), and so forth for higher-order states. Compared to the conventional point-like TLE scattering potential, a distinguishing feateure of the process here is that central to it is the formation of an effective non-local scattering potential mediated by the propagating bound states themselves. Inspired by giant atoms~\cite{FriskKockum2020, Gustafsson2014, Anton2014, Kockum2018, Kannan2020}, emitters coupled to a waveguide a multiple separate points, we formalize the concept of this nonlocal scattering potential as a pseudo-giant atom (PGA).

By utilizing real giant atoms (RGAs) with engineered coupling strengths and phases at multiple points in the waveguide, our cascaded architecture turns out to allow tuning of the $S \to D \to T$ conversion efficiency across nearly the full range. Furthermore, the scattering process exhibits remarkable unidirectionality rooted in quantum interference. Specifically, when the coupling parameters are tuned well, the single-photon (doublon) undergoes almost complete chiral conversion into a forward-propagating doublon (triplon), without residual single-photon transmission or reflection. The precise interference control within the giant atom's coupling profile enables an almost perfect cascaded flow of quantum states without backscattering losses, accompanied by sequentially increasing photon-number correlations.

Another defining feature of our platform is the automatic spatiotemporal sorting of the generated quantum state. Owing to the intrinsic photon-number-dependent group velocities ($v_g^{(1)}>v_g^{(2)}>v_g^{(3)}$) inherent to the distinct dispersion relations of the multi-photon state, the output wavepacket naturally separates into distinct spatially ordered temporal bins. The resulting output state is a programmable superposition of spatially isolated correlated quantum states:
\begin{equation}
\ket{\Psi_{\rm out}} = \alpha \ket{S}_{\tau_1} + \beta \ket{D}_{\tau_2} + \gamma \ket{T}_{\tau_3},
\end{equation}
where $\tau_1<\tau_2<\tau_3$ mark the arrival times of the single-photon, doublon, and triplon components, respectively, with the tunable conversion efficiencies $\alpha,\beta,\gamma \in [0,1)$ fulfilling $|\alpha|^2+|\beta|^2+|\gamma|^2=1$.

This work establishes cascaded inelastic scattering in nonlinear waveguides as a powerful paradigm for quantum state engineering. Unlike existing approaches~\cite{Mahmoodian2020, Tomm2023}, our architecture requires only a single TLE per scattering stage, thereby reducing system complexity while enhancing scalability and flexibility. The PGA framework extends scattering theory into the nonlinear waveguide regime by revealing how bound states generate emergent non-local scattering potentials. The combination of cascaded generation, programmable entanglement manipulation, and automatic spatiotemporal sorting of multi-photon states opens new possibilities for complex quantum many-body state synthesis, resource-efficient photonic quantum computing, and quantum networks requiring distributed entanglement. This unified approach bridges nonlinear quantum optics with correlated photonic matter, offering a versatile platform for both theoretical studies and technological applications.

This article is organized as follows. In \secref{sectionII}, we introduce the model of excited TLEs coupled to a nonlinear waveguide. We derive an effective non-local interaction Hamiltonian for this system and formalize the concept of the PGA, which captures the underlying multi-point scattering potential. In \secref{sectionIII}, we employ scattering theory to analytically obtain the output state when an incident single photon scatters off an excited TLE. We present numerical validation of these predictions and the PGA framework in \secref{sectionIIID}. In \secref{sectionIV}, we demonstrate the formation of spatially separated multi-photon wavepackets resulting from photon-number-dependent group velocities. Using an RGA configuration, we show full-range tunability of the conversion efficiency and unidirectional photon conversion. In \secref{high_order}, we present the cascaded inelastic scattering process, yielding a programmable superposition of spatiotemporally isolated correlated quantum states. We conclude with a summary and an outlook in \secref{sectionVI}.


\section{Model}
\label{sectionII}


\subsection{Excited two-level emitters coupled to a nonlinear waveguide}

We consider excited TLEs coupled to a nonlinear waveguide, consisting of an array of $N$ coupled nonlinear cavities, as shown in \figpanel{fig1}{a}. In the frame rotating with the cavity frequency $\omega _c$, the waveguide Hamiltonian is ($\hbar=1$ throughout this article)
\begin{equation}
H_w = - \sum_n \mleft[ J \mleft( a_n^\dag a_{n+1} + \text{H.c.} \mright) + \frac{U}{2} a_n^\dag a_n^\dag a_n a_n \mright],
\label{H_waveguide}
\end{equation}
where $a_n$ ($a_n^\dag$) is the bosonic annihilation (creation) operator for the $n$th cavity, $J$ is the hopping rate between adjacent cavities, and $U$ is the on-site Kerr nonlinear potential, which induces photon-photon interactions within each cavity.

In a linear waveguide ($U=0$), photons propagate freely without interaction. In the strong-interaction limit $U/J\rightarrow \infty$, however, photons undergo fermionization and obey an effective exclusion principle~\cite{Sakurai2017}. A finite nonlinear potential can lead to the formation of bound states in which multiple photons become localized, forming multi-photon quasiparticles~\cite{Winkler2006, Piil2007}, commonly referred to as boson stacks~\cite{Mansikkamaki2022}, close dimers~\cite{Valiente2008}, or photon molecules~\cite{Douglas2016}. For attractive interactions, these quasiparticle states lie below the uncorrelated photon bands, indicating effective binding. As illustrated in \figpanel{fig1}{b}, the energy spectrum of the nonlinear waveguide reveals a hierarchy of correlated few-photon states: two-photon (doublon) and three-photon (triplon) states emerge progressively below the uncorrelated single-particle band, forming a sequence of bound states caused by the photon-photon interaction~\cite{Mansikkamaki2022}. 

The single-photon states are described by the dispersion relation 
\begin{equation}
E_k = - 2 J \cos(k) ,
\end{equation}
with $k$ being the photon momentum. In the relative and center-of-mass coordinates, i.e., $r_c = n_1 - n_2$ and $x_c = (n_1 + n_2) / 2$, the dispersion relation of the doublon states is~\cite{Piil2007,Valiente2008}
\begin{equation}
E_K = - \sqrt{U^2 + \mleft[ 4 J \cos \mleft( K/2 \mright) \mright]^2} ,
\end{equation}
where $K$ denotes the center-of-mass momentum of the photon pair. 

For the multi-photon states, the on-site nonlinearity (at $r_c=0$) acts as an effective impurity, leading to an exponentially localized wave function in the relative coordinate. As a result, the two photons in the doublon state are bunched over a characteristic localization length $L_u(K)$ and propagate as a quasiparticle along the waveguide. The corresponding wave function takes the form
\begin{equation}
\Psi_K(x_c,r_c) = e^{iKx_c} u_K(r_c), \quad u_K(r_c) = u_0 e^{-|r_c| / L_u (K)} ,
\label{eq:DoublonWfct}
\end{equation}
where $u_0$ is a normalization constant~\cite{Piil2007, Wang2020, Wang2024}. In the three-photon subspace, a bound quasiparticle state (the triplon) emerges with dispersion relation $E_{\mathcal{K}}$, where $\mathcal{K}$ is the total momentum wave vector~\cite{Valiente2010,Compagno2017}. We denote the creation operators for single-photon, doublon, and triplon states as $a_{k}^{\dagger}$, $D_{K}^{\dagger}$, and $T_{\mathcal{K}}^{\dagger}$, respectively.

By adding the emitters and their coupling to the waveguide, still in the frame rotating with $\omega_c$, the total system Hamiltonian is
\begin{align}
H &= H_0 + H_{\rm int}, \label{H_total}
\\
H_0 &= H_w + \sum_i \frac{\Delta_{ei}}{2} \sigma_i^z , \label{eq:H0}
\\
H_{\rm int} &= \sum_i g_i \sigma_i^- a_{n_i}^\dag + \text{H.c.} ,
\end{align}
where $\sigma_{i}^{z}$ and $\sigma_{i}^{\pm}$ are the Pauli operators for the $i$th two-level emitter, $\Delta_{ei} = \omega_{ei} - \omega_c$ is the detuning of its transition frequency $\omega_{ei}$ from $\omega_c$, $n_i$ denotes its coupling site, and $g_i$ is its coupling strength. The excited and ground states of each emitter are denoted by $|e\rangle$ and $|g\rangle$, respectively.


\subsection{Scope of the study and initial conditions}

\begin{figure*}
	\centering \includegraphics[width = 0.9\linewidth]{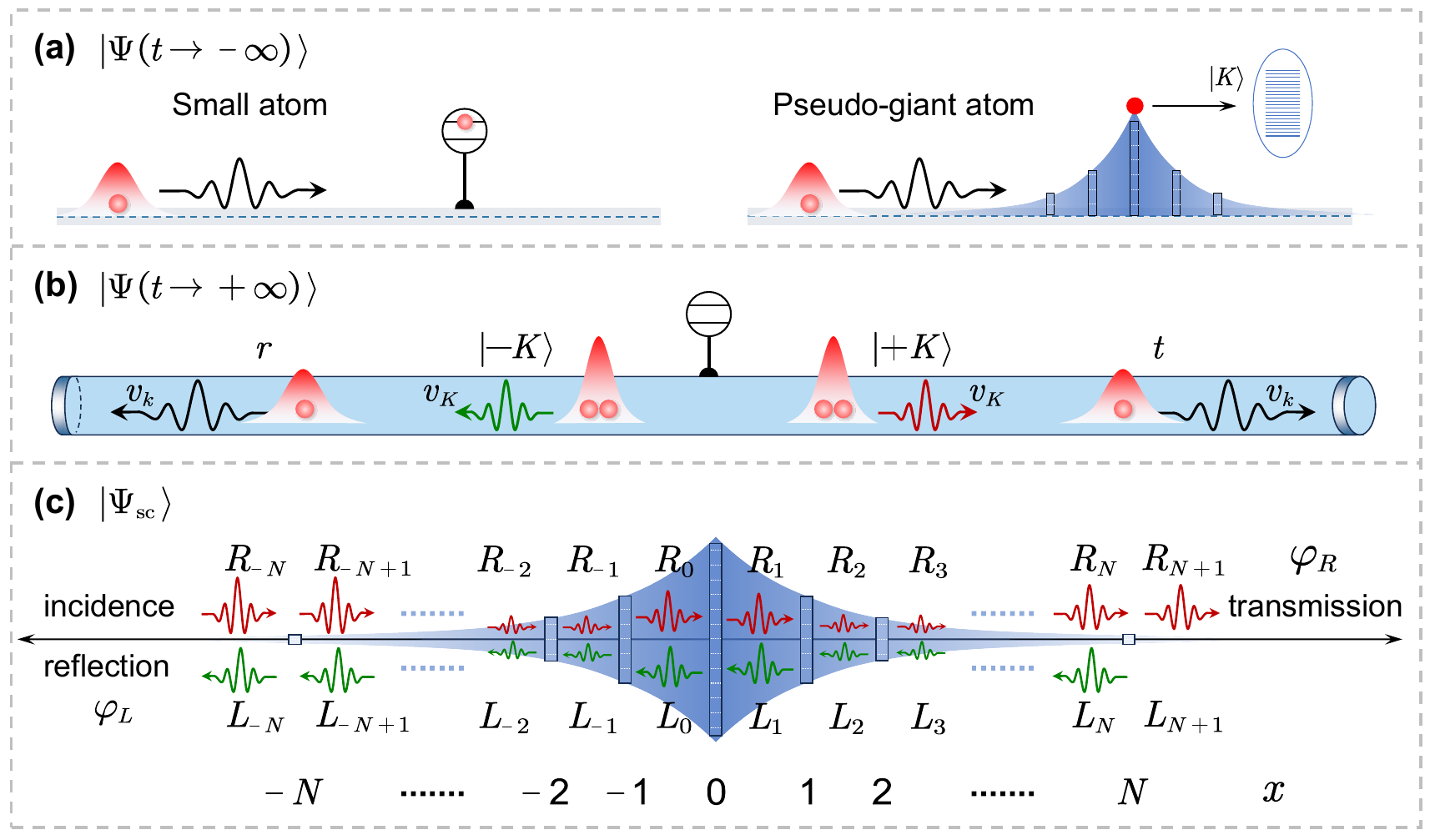}
	\caption{Schematic illustration of the scattering process at different stages.
	(a) Initial state at $t \to -\infty$. The left panel shows the physical setup in real space. The right panel depicts the effective ``pseudo-giant atom'' (PGA) description of the setup.
	(b) Final state at $t \to +\infty$, featuring the transmitted ($t$) and reflected ($r$) single-photon components, as well as right- and left-going doublon wavepackets $\ket{\pm K}$.
	(c) Scattering eigenstate structure, with right- and left-going single-photon wave functions denoted by $\varphi_{R,L}$. The labeled amplitudes indicate the spatial distribution of waves in different regions.}
	\label{fig2}
\end{figure*}

We consider a scattering scenario, in which an incident single-photon plane wave 
\begin{equation}
\ket{\psi} = \exp(i k_0 x)
\end{equation}
propagates from left to right and scatters off a far detuned, initially excited emitter located at $n_0$, as shown in \figpanel{fig2}{a}. In this setup, the nonlinearity of the waveguide enables an inelastic scattering process, in which the incoming single photon is converted into a correlated multi-photon state.

The two-photon state of the considered wQED system can be written as
\begin{align}
\label{total_state}
\ket{\Psi (t)} =& \sum_{n_1,n_2} c_\text{uc} (t) \ket{\psi_\text{uc} (n_1, n_2)}
\notag \\
&+ \sum_K c_K (t) \ket{K} + \sum_n c_{e,n} (t) \ket{e, n} ,
\end{align}
where $c_\text{uc}$ is the probability amplitude for finding two uncorrelated photons at sites $n_1$ and $n_2$, with $\ket{\psi_\text{uc} (n_1, n_2)}$ the corresponding wave function~\cite{Valiente2008}\footnote{This component is often referred to as the two-photon scattering state in the literature. To avoid confusion with the scattering states involved in the dynamical scattering process discussed later, we refer to it here as the uncorrelated state.}. The amplitude $c_{e,n}$ corresponds to the state with a single photon at site $n$ and the emitter being excited, while $c_K(t)$ represents the amplitude for the doublon mode $\ket{K}$.

In general, when the transition frequency of a TLE lies within the single-photon energy band, the TLE can resonantly exchange photons with the waveguide, enabling emission and absorption. In this regime, the TLE acts as a nonlinear impurity that blocks double occupancy, \textit{allowing at most one photon to occupy its site}. This effective on-site repulsion gives rise to few-photon bound states. For a detailed discussion of such bosonic impurity models, we refer the interested reader to Ref.~\cite{Shi2016}. Previous works have harnessed these correlated states to generate nonclassical light, typically by combining multiple emitters~\cite{Mahmoodian2020} or strong coupling schemes~\cite{Tomm2023} to overcome  the limited nonlinearity (i.e., shallow optical depth) inherent to a single TLE.

In our proposal, by contrast, the emitter is far detuned from the single-photon band, i.e., $|\Delta_e| \gg 2J$, which strongly suppresses resonant photon exchange. Both spontaneous and stimulated emission are inhibited. The uncorrelated scattering component $c_{\text{uc}}$ can be safely adiabatically eliminated, i.e., $c_{\text{uc}}\simeq 0$, and the emitter is frozen in its excited state, as shown in \figpanel{fig1}{c} at the initial moment. By neglecting the uncorrelated part, the Hilbert space is simplified as
\begin{equation}
\label{state_2}
\ket{\Psi (t)} \simeq \sum_K c_K (t) \ket{K} + \sum_n c_{e,n} (t) \ket{e, n} .
\end{equation}
Given $|\Delta_e| \gg 2J + g$, the emitter virtually participates in the formation of a bound state, with negligible population transfer to the ground state~\cite{John1990}. Consequently, the initial state is well approximated by
\begin{equation}
\ket{\Psi (t \rightarrow -\infty)} \simeq \sum_n e^{i k n} \ket{e, n} .
\end{equation}

Additionally, in the nonlinear bath, the dynamics of the emitter depend not only on the energy structure, but also on the photon state of the bath, owing to the photon-photon interactions~\cite{Imamoifmmode1997,Li2025}. In contrast to the linear regime, the nonlinear waveguide supports intrinsic multiphoton bound states, providing new channels for light-matter interaction. As shown in \figpanel{fig1}{b}, we consider a configuration in which the sum of the incident single-photon energy and the far detuned emitter transition frequency lies in the doublon band. Under this energy-matching condition, the incident photon unfreezes the emitter, enabling the joint creation of a two-photon bound state---a doublon.


\subsection{Interaction Hamiltonian: converting single-point interaction into a pseudo-giant atom}
\label{pseudo_giant_atom}

We now derive the effective interaction Hamiltonian between the $|e,n\rangle$ and $|K\rangle$ states. Starting from the general case, under the dipole approximation~\cite{Scully1997}, the emitter couples to the waveguide at a single site $n_0$, behaving as a point-like impurity [see the left panel of \figpanel{fig2}{a}]. The interaction Hamiltonian is then
\begin{equation}
\label{interaction_H_single}
H_{\rm int} = g \sigma ^- a_{n_0}^\dag + g \sigma ^+ a_{n_0} ,
\end{equation}
which describes the emission or absorption of a photon at $n_0$ via transitions between the emitter's excited $\ket{e}$ and ground $\ket{g}$ states~\cite{Scully1997}.

When the total energy of the photon and the emitter lies within the doublon band, a resonant process occurs: the simultaneous annihilation of the incident photon and decay of the emitter jointly excite a doublon mode $\ket{K}$. To describe this process, we project the interaction Hamiltonian in Eq.~\eqref{interaction_H_single} onto the reduced Hilbert space in Eq.~\eqref{state_2}. The matrix element for the transition $\ket{e, n} \to \ket{K}$ is~\cite{Wang2020, Wang2024, Li2025}
\begin{equation}
\langle K | \sigma^- a_{n_0}^\dag |e,n \rangle = \langle \text{vac}| D_K a_n^\dag a_{n_0}^\dag |\text{vac} \rangle = M^* (K, n, n_0) .
\end{equation}
Here, $M(K, n, n_0) \!=\! \langle \text{vac}|a_n a_{n_0} D_K^\dag|\text{vac} \rangle$ denotes the overlap between the doublon state and the two-photon state $\ket{n, n_0}$, given by [using Eq.~\eqref{eq:DoublonWfct}]
\begin{equation}
M (K, n, n_0) = \frac{\sqrt{2}}{\sqrt{N}} u_0 e^{i K (n + n_0) / 2} e^{- \frac{|n - n_0|}{L_u(K)}} .
\label{eq:M_transition}
\end{equation}
Recall that $L_u(K)$ is the correlated length of the doublon wave function in the relative coordinate $r_c = n - n_0$, characterizing the spatial extent over which the two photons are bound. The transition amplitude is significant only when the separation between the mobile photon (at $n$) and the emitter site ($n_0$) is smaller than $L_u(K)$, i.e.,  $|r_c| = |n - n_0| < L_u(K)$, ensuring sufficient wave function overlap. For simplicity and without loss of generality, we set $n_0 = 0$. The effective interaction Hamiltonian then becomes
\begin{align} 
H_{\rm int} &= \sum_{K, n} g M^* (K, n, 0) |K\rangle \langle e,n| + \text{H.c.}
\notag \\
&= \sum_{K, n} \frac{g_n}{\sqrt{N}} e^{- i K n/2} D_K^\dag \sigma_- a_n + \text{H.c.}
 \label{interaction_H_D}
\end{align}
where $g_n=\sqrt{2}gu_0\exp \left[ -|n|/L_u\mleft( K \mright) \right]$.

In contrast to the scattering process for a point-like emitter in linear waveguides~\cite{Fan2007, Fan2012, Sanchez2014, Shi2018}, this effective interaction Hamiltonian captures two distinct physical aspects, corresponding to the summations over the doublon momentum $K$ and the photon position $n$:
\begin{enumerate}
\item The emitter mediates a nonlinear scattering process that converts an incident single photon with wave vector $k$ into a propagating two-photon bound state $\ket{K}$. Unlike conventional scattering, the incident photon does not scatter off a single mode ($\ket{e}$ or $\ket{g}$), but off the entire available spectrum of correlated two-photon states.
\item Owing to the exponential localization of the doublon wave function, the transition amplitude is significant only when the incident photon is within a distance of the emitter $|n-n_0|<L_u(K)$. Thus, $L_u(K)$ sets the effective interaction range, defining the spatial scale over which the scattering process becomes appreciable. Consequently, the photon does not scatter off a point-like emitter, but off an extended, \textit{nonlocal} scattering potential.
\end{enumerate}

In Eq.~(\ref{interaction_H_single}), the emitter is modeled under the dipole approximation as having a point-like coupling to the waveguide, characterized by a local interaction strength $g$. However, in the effective Hamiltonian in Eq.~\eqref{interaction_H_D}, the scattering process is inherently nonlocal, involving contributions from multiple sites. The coupling strength at each site $n$ is weighted by the spatial profile of the doublon wave function, yielding an effective site-dependent amplitude $g_n$. This multi-point coupling structure resembles that of a \textit{giant atom}, which couples to the waveguide at multiple points~\cite{FriskKockum2020, Gustafsson2014, Anton2014, Kockum2018, Kannan2020, Wang2022}. Therefore, within the nonlinear scattering framework, the originally point-coupled TLE can be reinterpreted as a ``\textit{pseudo-giant atom}'' (PGA) with a spatially distributed coupling profile and a continuum of internal modes, as shown in the right panel of \figpanel{fig2}{a}.


\section{Nonlinear scattering theory with pseudo-giant atom formula}
\label{sectionIII}


\subsection{Introduction to the scattering process: general overview and theoretical framework}

To solve the scattering problem of a propagating single photon interacting with a TLE, we employ standard scattering theory based on the Lippmann--Schwinger formalism~\cite{Taylor2012, Fan2007}. Here, we provide a brief overview of the scattering process, as shown in \figpanel{fig1}{c}. As $t\rightarrow -\infty$, the incoming state $\ket{i}$ is a free state far from the interaction region and governed by $H_0$ [see Eq.~\eqref{eq:H0}]. Simultaneously, the emitter is highly detuned from the single-photon modes, suppressing spontaneous emission. 

Due to the nonlocal scattering potential, the scattering initiates at the outer edges of the PGA, where the interaction is weak but finite. As the incident photon propagates toward the center of the coupling region, the effective interaction strength increases gradually. The interaction then decreases symmetrically as the photon exits through the opposite edge of the PGA. In \figpanel{fig1}{c}, the shaded blue region depicts the non-local scattering potential, as governed by the interaction Hamiltonian in Eq.~\eqref{interaction_H_D}. During this process, part of the incident single photon hybridizes with the emitter to form a propagating doublon state, while the remainder is either transmitted or reflected. 

Consequently, at $t\rightarrow +\infty$, the final state $\ket{f_i}$ is a free state governed by $H_0$, far away from the interaction regime. Note that the final state contains not only single-photon, but also doublon components, as shown in \figpanel{fig2}{b}. The subscript $i$ indicates that the final state depends on the incoming state. 

The interaction Hamiltonian is non-local, yet its range is confined within the correlation length of the doublon wave function. Thus, the incoming photon interacts with the emitter only during a short  time interval before propagating away. Scattering theory remains applicable: the interaction can be treated as being adiabatically ``switched on'' as $t \to 0$, reaching its peak strength at $t=0$, and then being adiabatically ``switched off'' as $t\rightarrow+\infty$~\cite{Taylor2012,Fan2007}. 

We denote the interacting scattering state at $t=0$ by $\ket{\Psi_{\rm sc}}$. The three relevant states satisfy the Lippmann--Schwinger equations
\begin{align}
\ket{\Psi_{\rm sc}} & = \ket{i} + \frac{1}{E - H_0 + i 0^+} H_{\rm int} \ket{\Psi_{\rm sc}} , 
\label{LS_equation_i}
\\
\ket{\Psi_{\rm sc}} & = \ket{f_i} + \frac{1}{E - H_0 - i 0^+} H_{\rm int} \ket{\Psi_{\rm sc}} ,
\label{LS_equation_f}
\end{align}
where $(E - H_0 \pm i 0^+)^{-1} \equiv G_0^{R/A}$ are the free retarded and advanced Green's functions, respectively. These equations describe the time evolution of the system in complementary directions: Eq.~\eqref{LS_equation_i} corresponds to forward evolution from the initial state $\ket{i}$ to the scattering state $\ket{\Psi_{\rm sc}}$, mediated by $G_0^R$ under the adiabatic turn-on of the interaction; Eq.~\eqref{LS_equation_f} describes the backward evolution from the final state $\ket{f_i}$ to $\ket{\Psi_{\rm sc}}$, governed by $G_0^A$ during the adiabatic turn-off. Together, Eqs.~\eqref{LS_equation_i} and \eqref{LS_equation_f} imply the eigenvalue relations $H_0 \ket{i} = E \ket{i}$, $H \ket{\Psi_{\rm sc}} = E \ket{\Psi_{\rm sc}}$, and $H_0 \ket{f_i} = E \ket{f_i}$, where $E$ is the total energy of the system.

Note that the scattering state connects the incoming state to the final state, mediated by $G_0^{R/A}$. Given a specific incoming state, the energy is fixed by $H_0 \ket{i} = E \ket{i}$. The scattering state is then obtained via solving the secular equation $H \ket{\Psi_{\rm sc}} = E \ket{\Psi_{\rm sc}}$. The corresponding final state $\ket{f_i}$ follows directly from the Lippmann--Schwinger equation in Eq.~\eqref{LS_equation_f}.


\subsection{Scattering eigenstates}
\label{real_space_scattering}

To proceed, we determine the scattering state $\ket{\Psi_{\rm sc}}$ by 
solving the secular equation in real space. An equivalent formulation in 
momentum space, which yields identical results, is presented in the 
Appendix~\ref{momentum_space_derivation}. To characterize the propagation 
dynamics of single photons, we project the Hamiltonian onto the basis in 
Eq.~\eqref{state_2} and express it in continuous form via linearization 
around $k$~\cite{Fan2007,Shen2009_1,Shen2009_2}:
\begin{align}
&H_0 \simeq \int dx \, i v_k \mleft[ - c_R^\dag (x) \frac{\partial}{\partial x} c_R (x) + c_L^\dag (x) \frac{\partial}{\partial x} c_L (x) \mright]
\notag \\ 
& \qquad + \frac{\Delta_e}{2} \sigma _z + E_K D_K^\dag D_K, \label{real_H_1}
\end{align}
and the interaction Hamiltonian is
\begin{align}
H_{\rm int} = \int dx \, \biggl\{ & \sum_{K, n} g M^* (K, n, 0) \delta (x, n)
\notag \\
& D_K^\dag \sigma_- \mleft[ c_R (x) + c_L (x) \mright] + \text{H.c.} \biggr\}.
\label{real_H_int}
\end{align}
Here, $a_x \simeq c_R(x) + c_L(x)$, where $c_{R/L}$ are bosonic operators which create a right- or left-going photon at position $x$, and $v_k$ is the group velocity of the single photon. The Hamiltonian in Eq.~\eqref{real_H_1} conserves particle number, and the scattering state takes the form
\begin{align} 
\label{scattering_state}
\ket{\Psi_{\rm sc}} = & \int dx \mleft[ \varphi_R (x) c_R^\dag (x) + \varphi_L (x) c_L^\dag (x) \mright] \sigma_+ \ket{0} 
\notag \\
&+ \sum_K u_K D_K^\dag \ket{0} ,
\end{align}
where $\ket{0}$ is the vacuum state, $\varphi_{R,L}(x)$ is the right-/left-going single-photon wave function, and $u_K$ is the probability amplitude for the doublon state $\ket{K}$ to be excited. 

Within the PGA formalism, the interaction Hamiltonian [Eq.~\eqref{interaction_H_D}] only acts on integer lattice sites. This discrete coupling structure is preserved in Eq.~(\ref{real_H_int}), through the delta function $\delta(n,x)$. The summation over $n$ spans the nonlocal scattering region, anchoring the interaction to discrete spatial points. Consequently, the wave function varies only at these points, resulting in piecewise-constant behavior. Specifically, the single-photon wave function remains constant between adjacent lattice sites and changes only at integer positions $x=n$:
\begin{widetext}
\begin{align}
\varphi_R(x) & = e^{+ i k x} \mleft\{ R_{-N} \Theta [- (x + N)] + \sum_{n = - N + 1}^N R_n \mleft[ \Theta (x - n) - \Theta (x - n + 1) \mright] + R_{N+1} \Theta (x - N) \mright\}, 
\label{phi_R}
\\
\varphi_L(x) & = e^{- i k x} \mleft\{ L_{-N} \Theta [- (x+N) ] + \sum_{n = - N + 1}^N L_n \mleft[ \Theta (x - n) - \Theta (x - n + 1) \mright] + L_{N+1} \Theta (x - N) \mright\} , 
\label{phi_L}
\end{align}
\end{widetext}
where $\Theta (x)$ is the Heaviside step function. This ansatz is analogous to a scattering process in which a single photon scatters off a \textit{real giant atom} with multiple coupling points~\cite{Wang2019, Chen2022, Xue2025}. The coefficients $R_n$ and $L_n$ denote the probability amplitudes of the wave function in the respective regions, as shown in \figpanel{fig2}{c}. 

Returning to Eq.~(\ref{LS_equation_i}), under the assumption of an initial plane wave, the fore and aft of the wave function are given by
\begin{gather}
R_{-N} = 1, \quad R_{N+1} = t \quad L_{-N} = r, \quad L_{N+1} = 0 ,
\end{gather}
where the right-going wave function is assumed to be $1$ at the position $x < N$, and $t$ and $r$ are the transmission and reflection amplitudes, respectively. The initial state gives the total energy of the system: $E = \Delta_e + \omega_{k_0} \simeq \Delta_e + v_{k_0} k_0$. Substituting the Hamiltonian in Eq.~\eqref{real_H_1} and the scattering-state ansatz in Eq.~\eqref{scattering_state} into the secular equation $H \ket{\psi} = E \ket{\psi}$, we derive the equations
\begin{align}
& \mleft( E_{K_r} - E_K \mright) u_K =
\notag \\
& \frac{1}{\sqrt{N}} \sum_n g_n e^{- i K n / 2} \mleft[ \mathcal{P}_R^+ (n) + \mathcal{P}_L^+ (n) \mright] / 2 ,
\label{motion_equation_u_K}
\\
& i v_k \mathcal{P}_R^- (n) = i v_k \mathcal{P}_L^- (n) = \frac{1}{\sqrt{N}} \sum_K g_n e^{i K n / 2} u_K , \label{motion_equation_R_L}
\end{align}
where $E_{K_r} = \mleft( \Delta_e + \omega_{k_0} \mright)$. Here, $K_r$ denotes the resonant doublon mode, and
\begin{align} 
\mathcal{P}_R^\pm (n) &= \mleft( R_{n+1} \pm R_n \mright) e^{+ i k n} , 
\label{R_pm}
\\
\mathcal{P}_L^\pm (n) &= \mleft( L_n \pm L_{n+1} \mright) e^{- i k n} .
\label{L_pm}
\end{align}

Equations~\eqref{motion_equation_u_K} and~\eqref{motion_equation_R_L} reveal that the single-photon mode and the doublon mode are coupled, via the far detuned emitter. First, Eq.~\eqref{motion_equation_u_K} describes the conversion of a single photon into a doublon through the pseudo-coupling points at \textit{integer lattice sites}. The effective coupling strength is proportional to $g_n\exp(-iKn/2)$ and depends on the single-photon probability amplitude at these discontinuities. In contrast, Eq.~\eqref{motion_equation_R_L} governs the reverse process---the decay of a doublon back into a single photon via the same pseudo-coupling structure.

Under the initial condition, Eq.~\eqref{motion_equation_u_K} is transformed into
\begin{gather} 
\label{origin_u_PM}
u_K = \frac{1}{\sqrt{N}} \sum_n \frac{g_n e^{- i K n / 2}}{E_{K_r} - E_K + i 0^+} \frac{\mathcal{P}_R^+ (n) + \mathcal{P}_L^+ (n)}{2} .
\end{gather}
The primary task is now to determine the functional forms of $\mathcal{P}_{R/L}^\pm (x)$. Substituting Eq.~\eqref{origin_u_PM} into Eq.~\eqref{motion_equation_R_L} eliminates $u_K$ and yields a closed equation for the single-photon wave function:
\begin{align}
& \mathcal{P}_{R/L}^- (n) = 
\notag \\ 
& - \frac{1}{v_k v_{K_r}} \sum_{n'} g_n g_{n'} e^{i K_r |n - n'| / 2} \mleft[ \mathcal{P}_R^+ (n') + \mathcal{P}_L^+ (n') \mright] / 2 .
\label{P_x}
\end{align}
These equations (for arbitrary $n$) exhibit a convolution structure, arising from the multipoint coupling of the PGA, which acts as an effective nonlocal scattering potential. The scattering response at any given point depends on the cumulative effect across the entire PGA region, rather than on local properties alone. Crucially, this spatially distributed pseudo-coupling gives rise to complex interference effects, discussed in the following section.

In a conventional system with light--matter interaction, an emitter with discrete energy levels couples to a bath with continuous modes, and the dynamics are governed primarily by the bath's density of states, as described by Fermi's golden rule~\cite{Scully1997}. In contrast, in the present system, the emitter mediates the transition between two quantum fields, the single-photon and the doublon modes. Consequently, the coefficients in Eq.~(\ref{P_x}) depend on the densities of both continua, $\rho_k=1/v_{k_0}$ and $\rho_K=1/v_{K_r}$.

Note that in Eq.~\eqref{P_x}, the number of equations matches the number of unknown variables. By solving for $\mathcal{P}_{R/L}^\pm (n)$ at $k = k_0$, we determine their values and subsequently obtain the amplitudes of the single-photon wave function ($R_n, t, L_n, r$) via Eqs.~\eqref{R_pm} and \eqref{L_pm}. Substituting $\mathcal{P}^\pm_{R/L}(n)$ back into Eq.~\eqref{motion_equation_u_K}, the doublon amplitude $u_k$ can also be obtained. The solution framework, including the wave-function ansatz [Eqs.~\eqref{phi_R} and~\eqref{phi_L}] and the auxiliary variables $\mathcal{P}_{R/L}^\pm(n)$ [Eqs.~\eqref{R_pm} and~\eqref{L_pm}] can also be applied to the case of a real giant atom (RGA). The key distinction lies in the nature of the scattering potential: in the RGA, it arises from discrete, localized energy levels, whereas in PGA, it stems from continuous, propagating doublon modes.

The overall summation structure reflects the occurrence of multiple scattering events and the interference between distinct scattering pathways. At its core, this equation describes the propagation and interference of quantum waves in the presence of a nonlocal potential. It further  reveals a fundamental connection between scattering theory and the convolution structure. This nonlocal scattering potential is central to the nonlinear scattering behavior observed in this system.


\subsection{The output field for the single-photon plane-wave input}

The final state takes the form
\begin{align}
\ket{f_i} \simeq & \int dx \mleft[ f_R (x) \ket{e, x_R} + f_L (x) \ket{e, x_L} \mright]
\notag \\
& + \sum_{\pm K} f_{\pm K} \ket{\pm K} ,
\end{align}
where the doublon mode is separated into positive and negative components to distinguish right- and left-going waves. We have so far obtained the scattering state for a single-photon plane-wave input. To obtain the output state, we return to Eq.~\eqref{LS_equation_f} and project the final state onto $\ket{e, x_{R/L}}$ and $\ket{K}$, respectively. Here, $f_{R,L,\pm K}$ denote the probability amplitudes of right-/left-going single-photon and doublon modes in the final state, as shown in \figpanel{fig2}{b}.


\subsubsection{Doublon component}

Substituting the interaction Hamiltonian in Eq.~\eqref{real_H_int} and the scattering state in Eq.~\eqref{scattering_state} into Eq.~\eqref{LS_equation_f}, and projecting onto the $\ket{+K}$ state, we obtain
\begin{align}
& \braket{+K | f} 
\notag \\
& = \braket{+K | \Psi_{\rm sc}} - \bra{+K} \frac{1}{E_{K_r} - H_0 - i 0^+} H_{\rm int} \ket{\Psi_{\rm sc}} 
\label{f_k_real_space_1} \\
& = \frac{1}{\sqrt{N}} \sum_x \sum_{K'} g_n \mathcal{P}_{R+L}^+ (n)  
\notag \\
& \bra{+K} \frac{e^{- i K' n / 2}}{E_{K_r} - E_{K'} + i 0^+} - \frac{e^{- i K' n / 2}}{E_{K_r} - E_{K'} - i 0^+} \ket{K'} .
\label{f_k_real_space_2}
\end{align}
The two Green's function integrals are~\cite{Taylor2012}
\begin{align}
& \frac{1}{N} \sum_K \frac{e^{- i K n / 2}}{E_{K_r} - E_K + i 0^+} \ket{K} = - \frac{i}{v_{K_r}} \times
\notag \\
& \mleft[ e^{- i K_r n / 2} \Theta (-n) \ket{+K_r} + e^{+ i K_r n / 2} \Theta (+n) \ket{-K_r} \mright], 
\label{Green_retarded}
\\
& \frac{1}{N} \sum_K \frac{e^{- i K n / 2}}{E_{K_r} - E_K - i 0^+} \ket{K} = + \frac{i}{v_{K_r}} \times
\notag \\
& \mleft[e^{- i K_r n / 2} \Theta (+n) \ket{+K_r} + e^{+ i K_r n / 2} \Theta (-n) \ket{-K_r} \mright] .
\label{Green_advanced}
\end{align}
In the integrals, we apply the Wigner--Weisskopf approximation, in which $E_K$ is linearized around $K = K_r$~\cite{Scully1997}.

The resulting equations provide an intuitive physical picture. The first term in Eq.~\eqref{f_k_real_space_1} describes the evolution of the incident single photon from $t = - \infty$ to $t = 0$, governed by the retarded Green's function. The propagator $G_0^R$ yields a contribution from the nonlocal scattering potential in the region $x < 0$, represented by $\Theta(-x)$ in Eq.~\eqref{Green_retarded}, as the wave propagates forward in time from $x = - \infty$ to $x = 0$. The second term describes evolution from $t = \infty$ to $t = 0$, governed by the advanced Green's function $G_0^A$. In this case, $G_0^A$ accounts for the influence of the scattering potential in the region $x > 0$, captured by $\Theta(+x)$ in Eq.~\eqref{Green_advanced}, as the wave evolves backward in time from $x = + \infty$ to $x = 0$. Combining both spatial directions, the initial state evolves under the complete nonlocal potential. 

The final state is then given by
\begin{align}
& f_{+K_r} = \langle +K_r | f\rangle
\notag \\
& = - \frac{i}{v_{K_r}} \sum_n g_n \mathcal{P}_{R+L}^+ (n) \mleft[ \Theta (-n) + \Theta (+n) \mright] e^{- i K_r x / 2}
\notag \\
& = - \frac{i}{v_{K_r}} \sum_n g_n \mathcal{P}_{R+L}^+ (n) e^{- i K_r n / 2} .
\label{real_space_final_state_pK}
\end{align}
To make the expression more concise, we here defined
\begin{equation}
\mathcal{P}_{R+L}^+ (n) = \frac{\mathcal{P}_R^+ (n) + \mathcal{P}_L^+ (n)}{2} .
\end{equation}
For the left-going doublon mode $-K$, the first (second) term yields $\Theta(+n)$ [$\Theta(-n)$]:
\begin{equation}
\label{real_space_final_state_mK}
f_{-K_r} = \braket{-K_r | f} = - \frac{i}{v_{K_r}} \sum_n g_n \mathcal{P}_{R+L}^+ (n) e^{+ i K_r n / 2}.
\end{equation}

Up to now, we obtained the probability amplitudes for the $\pm K_r$ doublon modes. Note that, for the input single-photon state $\ket{i}$, it is straightforward to show that
\begin{equation}
\braket{\pm K | i} = \braket{\pm K | \Psi_{\rm sc}} - \bra{\pm K} \frac{1}{E_{K_r} - H_0 + i 0^+} H_{\rm int}  \ket{\Psi_{\rm sc}}.
\end{equation}
The first term is identical to the $\ket{f_i}$ state, while the second term takes the retarded form $1/(E_{K_r}-E_{K}+i0^+)$, which coincides with the first, i.e., 
\begin{equation}
\label{real_space_initial_state_K}
\braket{\pm K | i}=0. 
\end{equation}
The process can be seen as the evolution from $t=-\infty$ to $t=0$, followed by a return to $t=-\infty$, effectively reconstructing the initial state.


\subsubsection{Single-photon component}

Next, we derive the single-photon transmission and reflection amplitudes by projecting onto $\ket{e,x_{R/L}}$:
\begin{align}
& f_R = \braket{e, x_R | f}
\notag \\
& = \braket{e, x_R | \Psi_{\rm sc}} - \bra{e, x_R} \frac{1}{E_{K_r} - H_0 - i 0^+} H_{\rm int} \ket{\Psi_{\rm sc}} 
\notag \\
& = \varphi_R (x) -
\notag \\
& \int dx' \sum_n i v_k \mathcal{P}_R^- (n) \delta (x' - n) \bra{x_R} \frac{1}{E_{K_r} - H_0 - i 0^+} \ket{x'_R} 
\notag \\
& = \varphi_R (x) - \frac{1}{v_{k_0}} \sum_n \mathcal{P}_R^- (n) e^{+ i k_0 (x - n)} \Theta (n - x),
\label{f_project_onto_x}
\end{align}
and $f_L$ can be similarly obtained as
\begin{equation}
f_L = \varphi_L (x) - \frac{1}{v_{k_0}} \sum_n \mathcal{P}_L^- (n) e^{- i k_0 (x - n)} \Theta (x - n) .
\label{f_project_onto_x_l}
\end{equation}
During the derivation, we employ the Fourier transform to obtain the Green's function integral
\begin{align}
& \bra{x} \frac{1}{E_{K_r} - H_0 - i 0^+} \ket{x'} = \frac{1}{2\pi} \int dk \frac{e^{i k (x - x')}}{\omega_{k_0} - \omega_k - i 0^+}
\notag \\
& = \begin{cases}
(i / v_{k_0}) e^{+ i k_0 (x - x')} \Theta (x' - x) |_{+k_0} \\
(i / v_{k_0}) e^{- i k_0 (x - x')} \Theta (x - x') |_{-k_0} 
\end{cases} .
\end{align} 

Note that this is the advanced Green's function, describing evolution from $t = + \infty$ to $t = 0$. Consequently, the response at position $x$ due to the excitation at $x'$ is mediated by the right-going mode $+k$ when $x' > x$, and by the left-going mode $-k$ when $x' < x$. By substituting the wave-function ansatz [Eqs.~\eqref{phi_R} and \eqref{phi_L}] into the expressions for the final state [Eqs.~\eqref{f_project_onto_x} and \eqref{f_project_onto_x_l}], and carefully analyzing the solution in each piecewise region, we obtain the single-photon components of the initial and final states:
\begin{align}
\braket{e, x | i} &= e^{i k_0 x},
\label{real_space_initial_state_x}
\\
\braket{e, x | f_i} &= t e^{i k_0 x} + r e^{- i k_0 x} , 
\label{real_space_final_state_x}
\end{align}
where $t = f_R$ and $r = f_L$ denote the transition and reflection 
amplitudes, respectively, of the single photon. 

Equations~(\ref{real_space_initial_state_K}) and (\ref{real_space_initial_state_x}) confirm that the ansatz [Eq.~\eqref{scattering_state}] is a scattering eigenstate under the given initial condition. Ultimately, we obtain the final state, described by Eqs.~(\ref{real_space_final_state_pK}), (\ref{real_space_final_state_mK}), and (\ref{real_space_final_state_x}), for an incident single-photon plane wave $\exp(ik_0x)$.


\subsubsection{Flux conservation}

In the waveguide, the energy carried by a wave packet propagates at the group 
velocity~\cite{brillouin2013wave,Bers2000}. The energy flux $j$ can be 
approximated as the product of the amplitude of the wave packet and its group 
velocity, i.e., $j =  P v_g$. In conventional wQED, most studies focus on 
single-photon scattering, where both the incident and final states contain 
only single-photon components. This leads to the familiar flux-conservation 
relation $t^2+r^2=1$~\cite{Fan2007,Sanchez2014}. However, in our proposal, 
the single photon can be inelastically scattered into a doublon mode, with a 
different group velocity, i.e., $v_k\ne v_K$. As a result, flux conservation 
must account for contributions from both single-photon and doublon 
excitations. Now, the conservation law takes the form
\begin{align}
v_k P_S^i + v_K P_D^i = v_k P_S^f + v_K P_D^f,
\\
v_k = v_k \mleft( t^2 + r^2 \mright) + v_K \sum_{\pm} \mleft| u_{\pm K_r} 
\mright|^2,
\label{flux_conservation}
\end{align}
where $P^{i/f}_{S/D}$ denote the single-photon and doublon population in the 
initial and final state, respectively. We refer interested readers to 
Ref.~\cite{Peres2009} for a more detailed discussion of probability currents 
in systems with position-dependent velocity.

The incident wave packet is assumed to be a single photon with unit 
population in our numerical simulations. Consequently, in order to satisfy 
flux conservation, the final state is written as~\cite{Peres2009,Sakurai2017}
\begin{align}
\ket{f_i} = & \sum_x  \mleft( t e^{+ i k_0 x} + r e^{- i k_0 x} \mright) \ket{e, x} 
\notag \\
& + \sqrt{\frac{v_{K_r}}{v_{k_0}}} u_{\pm K_r} \ket{\pm K_r} .
\label{revised_f}
\end{align}
The group-velocity factor arises from the distinct dispersion relations ($E_k$, $E_K$), and serves as a critical precursor to spatiotemporal photon-number-dependent sorting.


\subsection{Multi-point scattering and interference within the PGA framework}
\label{sectionIIID}

We consider an incident wave packet centered at momentum $k_0$ with a spectral width $L_G$. Analogous to spontaneous emission and under the Wigner--Weisskopf approximation, i.e., $L_G \ll 4J$, the final state is dominated by spectral components near $k_0$, and remains insensitive to the temporal envelope of the input, as further discussed in Appendix~\ref{Appendix_B}. This behavior contrasts fundamentally with stimulated emission from an initially excited TLE, where the output field depends on the spectral and temporal profile of the input due to the finite emission linewidth $\Gamma$~\cite{Fan2012}. 

To validate the PGA formula, we numerically simulate the evolution of a Gaussian wave packet governed by the full Hamiltonian [Eq.~\eqref{H_total}]. The incident wave packet is given by
\begin{equation}
\psi_G (k) = \mleft( \frac{1}{2 \pi L_G^2} \mright)^{\frac{1}{4}} \exp \mleft[ - \frac{(k - k_0)^2}{4 L_G^2} \mright] . 
\label{Gaussian_wave}
\end{equation}
Its Fourier transform yields a spatially localized single-photon state centered at $x_0$. In the limit $L_G \to 0$ (narrow momentum spread), it approximates a plane wave.

\begin{figure}
	\centering \includegraphics[width = \linewidth]{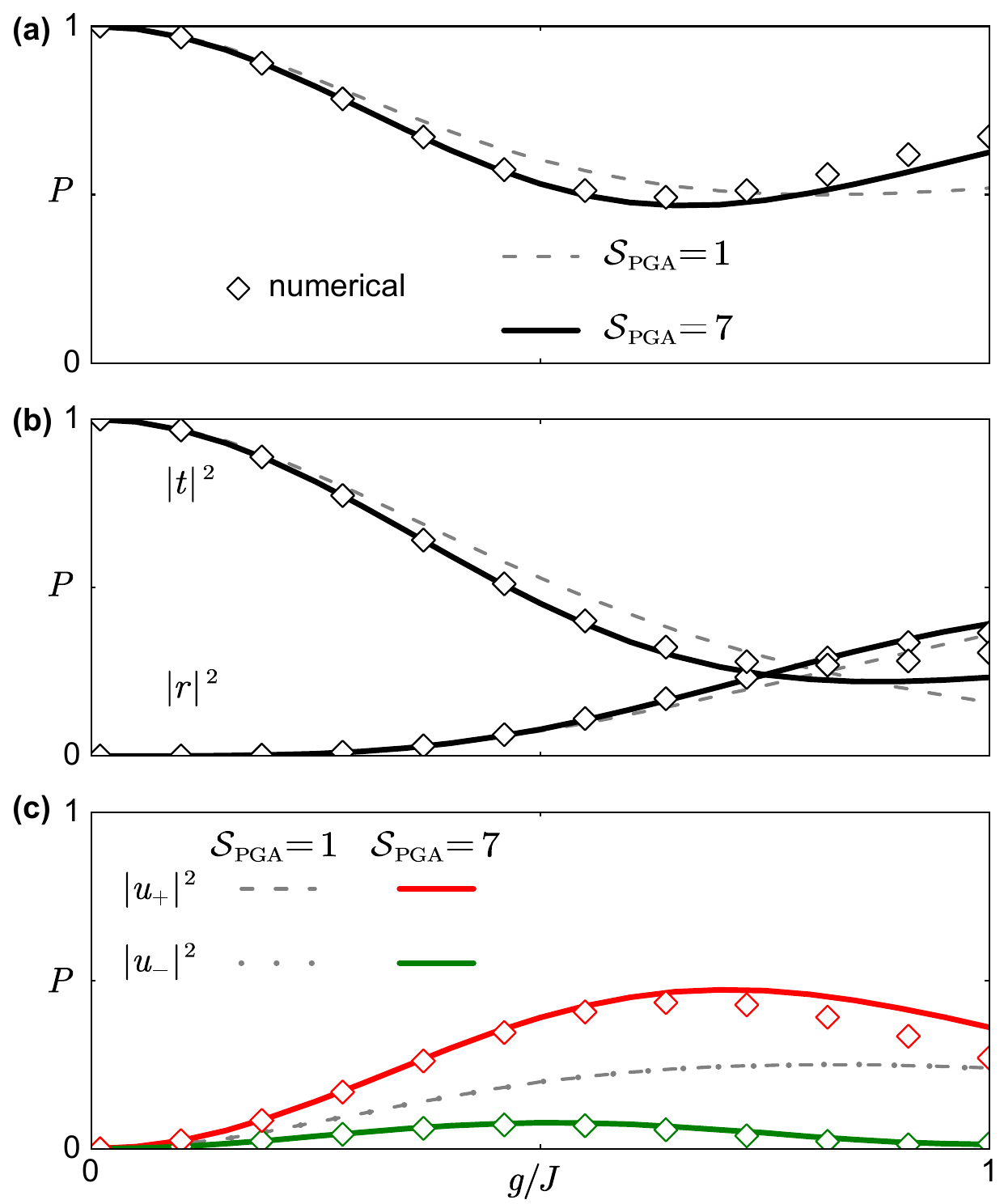}
	\caption{Populations of 
	(a) the emitter excitation, 
	(b) single-photon transmission $|t|^2$ and reflection $|r|^2$, and 
	(c) right- and left-going doublons $|u_{\pm}|^2$, 
	as functions of the coupling strength $g$. Numerical results are indicated by diamond markers. Solid colored curves are obtained by solving Eq.~\eqref{revised_f} with the size of the PGA $\mathcal{S}_\text{PGA} = 7$, while thin gray dashed curves represent the pointlike scenario $\mathcal{S}_\text{PGA} = 1$. Parameters are $L_G = 0.002$, $U = 6J$, $\Delta_e = -6.633J$, and $k_0 = \pi / 2$.}
	\label{fig3}
\end{figure}

We set the waveguide length to $N=2000$, resulting in a Hilbert-space dimension of $N_H \simeq N^2/2 = 2 \times 10^6$. Figure~\ref{fig3} displays the populations of key physical quantities as functions of $g$. These quantities include the emitter excitation, single-photon transmission $|t|^2$ and reflection $|r|^2$, and right- and left-going doublon populations $|u_{\pm}|^2$. Owing to the strong binding of the doublon, the numerical simulation results are defined as
\begin{equation}
|t|^2 / |r|^2 = \sum_{n \neq 0} |c_{e,n}|^2 , \quad |u_{\pm}|^2 = \sum_{x_c \neq 0, r = 0, 1, 2} |c_{x_c, r}|^2 .
\end{equation}
The two-photon state is dominated by the doublon component, with negligible contributions from uncorrelated photon pairs, thereby justifying the approximation $c_\text{uc} \simeq 0$.

\begin{figure}
	\centering \includegraphics[width = \linewidth]{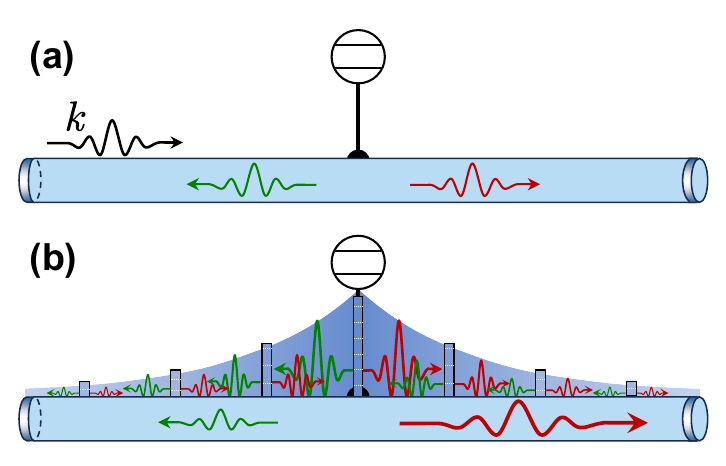}
	\caption{Schematic illustration of interference in doublon generation.
	(a) The pointlike scenario with equal distribution doublons in the right- and left-going modes.
	(b) PGA scenario: interference among seven pseudo-coupling points leads to an asymmetric doublon distribution.}
	\label{fig4}
\end{figure}

The analytical results are given by Eq.~\eqref{revised_f}. We first consider the ``pointlike'' scenario, in which only the local term $\delta (n, n_0) M (K, n, n_0)$ is retained [the size of the PGA $\mathcal{S}_\text{PGA} = 1$; see \figpanel{fig4}{a}], corresponding to the gray dashed curves in Fig.~\ref{fig3}. Within this approximation, the doublon distribution is symmetric, with all doublons formed at $r_c = 0$~\cite{Sanchez2014}. However, as shown in \figpanel{fig3}{c}, the numerical results exhibit a distinctly asymmetric doublon distribution. This clear discrepancy demonstrates that the pointlike approximation fails to capture the essential physics of this process.

Therefore, to obtain an accurate analytical description of this nonlinear scattering process, it is essential to incorporate \textit{the effective pseudo-multiple coupling points of the PGA}. Due to the exponential spatial decay $\exp[- |r_c| / L_u (K)]$, the transition amplitude $M (K_r, n, n_0)$ becomes negligible for $r_c = |n - n_0| > L_u (K)$. Specifically, we truncate the range of PGA at $r_c=3$, retaining seven pseudo-coupling points at $n=0,\pm 1, \pm 2,\pm3$ [i.e., $\mathcal{S}_\text{PGA} = 7$; see \figpanel{fig4}{b}]. By solving the coupled system of equations in Eq.~\eqref{P_x} and following the analytical derivation, we obtain the final state given by Eq.~(\ref{revised_f}). Figure~\ref{fig3} shows the agreement between the analytical results (solid colored curves) and full dynamical simulations (diamonds), which validates the PGA framework. Crucially, at each pseudo-coupling point, the doublon is generated symmetrically in both directions with amplitudes and phases; see \figpanel{fig4}{b}. Analogous to the multipoint interference in real giant atoms~\cite{Wang2022}, the doublons generated at different spatial positions interfere coherently, which results in an asymmetric doublon distribution between the left- and right-going directions, i.e., $|u_{+K_r}|^2 \neq |u_{-K_r}|^2$. 

In this physically local single-point coupling scenario, the nonlocal doublon wave function underpins effective multipoint scattering dynamics within the PGA framework, enabling coherent interference and leading to directional asymmetry. Although multiple interference pathways coexist, the effective coupling strength at each pseudo-site scales linearly with the physical coupling $g_n \propto g$, thereby limiting the degree of tunability. As a result, scattering is dominated by the central channel at $n = 0$, since $g_{n=0} \gg g_{n \neq 0}$. As shown in Fig.~\ref{fig3}, the conversion efficiency reaches a maximum, yet a finite doublon population persists in the incident direction. Thus, achieving flexible modulation of doublon generation and realizing optimal unidirectional cascaded scattering remain challenges.


\section{Spatially separated photonic wave packets sorted by photon number}
\label{sectionIV}

\begin{figure*}
	\centering \includegraphics[width = \linewidth]{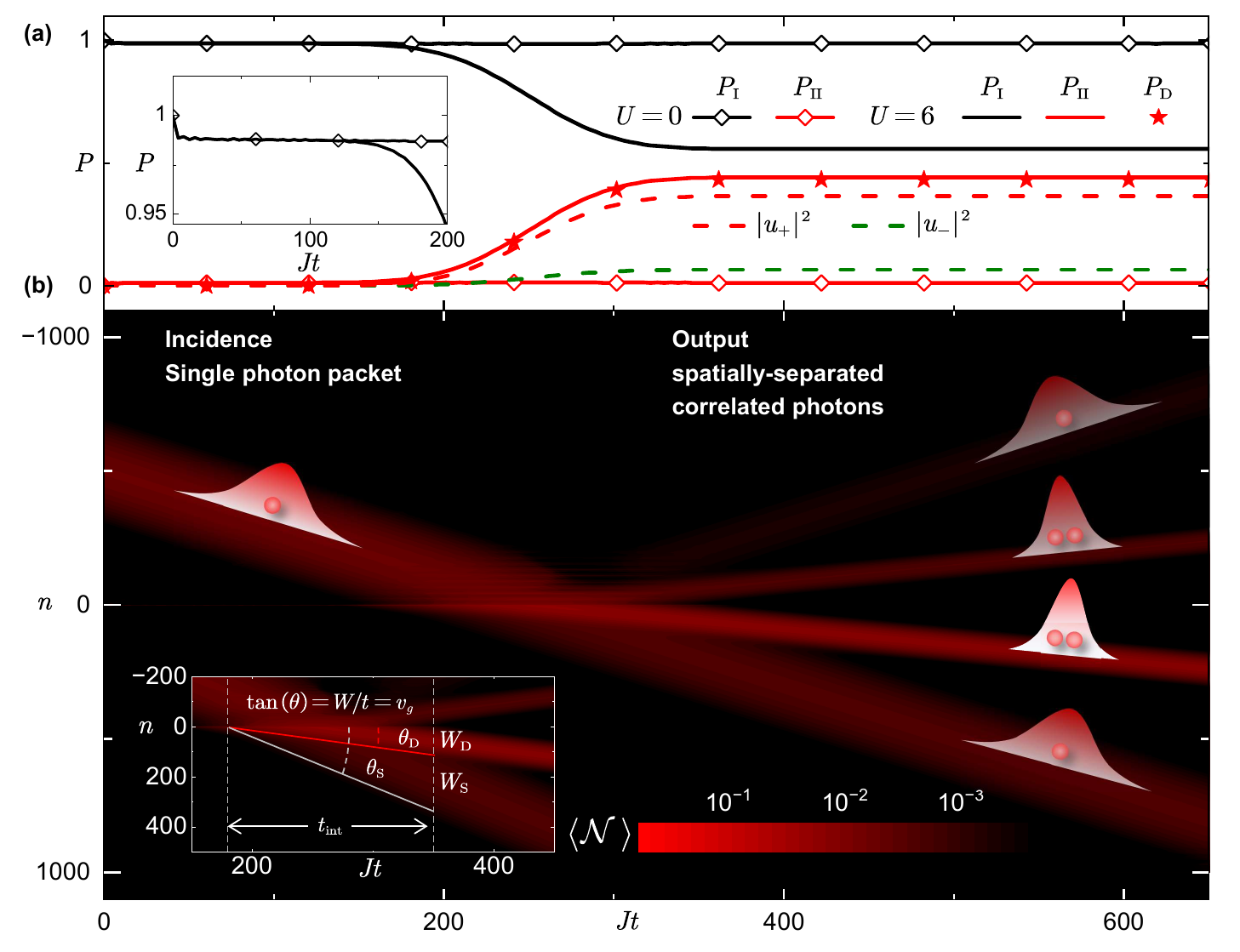}
	\caption{Scattering into spatially separated correlated photons.
	(a) Time evolution of the populations, including the single-photon state $P_{\rm I}$, the full two-photon state $P_{\rm II}$, the doublon state $P_D$, and the right-/left-going doublon components $|u_{\pm}|^2$, for different nonlinear strengths $U = 0$ and $U = 6 J$. The inset shows the initial bound state population.
	(b) Photon number $\langle \mathcal{N} \rangle $ as a function of position $n$ and time $J t$. Following scattering, the incident photon splits into four distinct components: transmitted ($t$) and reflected $(r)$ single photons, and right- and left-going doublon states $u_\pm$. The inset displays the natural spatial separation between single-photon and doublon states, arising from their distinct group velocities $v_{k_0}$ and $v_{K_r}$, respectively. The coupling strength is here set to $g = 0.5J$; all other parameters are the same as in Fig.~\ref{fig3}. }
	\label{fig5}
\end{figure*}

To further investigate the scattering dynamics, we show in \figpanel{fig5}{a} the time evolution of key state populations for $U = 0J$ and $6J$, including: the single-photon probability $P_{\rm I} =\sum_n |c_{e, n}|^2$; the doublon population $P_D = \sum_{x_c, r_c = 0, 1, 2} |c_{x_c, r}|^2$ (restricted to tightly bound states with $r_c \leq 2$); and the total two-photon population $P_{\rm II} = \sum_{n_1, n_2} |c_{n_1, n_2}|^2$, which encompasses both doublon and uncorrelated components. For $U = 0$, the incident single photon passes freely by the emitter without interaction and no doublon states are formed, leaving $P_{\rm I} \simeq 1$ and $P_{\rm II} = |c_\text{uc}|^2 \simeq 0$ throughout the evolution. In contrast, for $U = 6J$, the single photon cooperates with the emitter to jointly form a doublon state. Importantly, comparison between the $U = 0$ and $U = 6J$ cases reveals that over the entire process, the excitation into uncorrelated states is negligible, and the two-photon component is dominated by the doublon, with $P_{\rm II} \simeq P_D$.


\subsection{Spatial separation of scattered wavepackets}

To characterize the photon field, \figpanel{fig5}{b} shows the expectation value of the photon number as a function of position $n$ and time $t$, defined by
\begin{equation}
\langle \mathcal{N} \rangle = \bra{\Psi(t)} a_n^\dag a_n \ket{\Psi(t)} ,
\end{equation}
where $|\Psi(t) \rangle$ is the total two-photon state given by Eq.~\eqref{total_state}. As the incident single-photon packet scatters off the emitter, the outgoing photonic field splits into four distinct components: the transmitted $t$ and reflected $r$ single-photon components, and the right- and left-going doublon components $|u_{\pm K_r}|^2$, corresponding to the final state in Eq.~\eqref{revised_f}. \Figurepanel{fig5}{b} illustrates \textit{the natural spatial separation between single-photon and doublon wave packets within the same waveguide.} This separation arises because the doublons, as bound two-photon states, propagate at a slower group velocity compared to the free single photons. The disparity in group velocity directly leads to their spatial divergence.

In Ref.~\cite{Mahmoodian2020}, the authors achieve spatial separation of multiphoton states through a photon-number-dependent Wigner delay, arising from resonant photon exchange of a TLE. However, owing to the limited optical depth of a single TLE, the induced time delay is small and scales as $\tau_n=4/(n^2\Gamma)$ for an $n$-photon state traversing the TLE. Consequently, significant temporal delays require either multiple emitters arranged in an ordered array or operation in the strong coupling regime, both of which increase experimental complexity. By contrast, in our proposal, the nonlinearity is intrinsic to the waveguide bath via the interaction strength $U$, making the multi-photon bound states inherent to the bath. A single far detuned emitter couples to the nonlinear waveguide and mediates the scattering of an incident single photon into a two-photon bound state, with possible extension to higher-order correlated photon states (see \secref{high_order}). Owing to the distinct group velocities of the scattered single-photon and bound-state components, these quantum excitations naturally separate in space after the interaction, without the need for engineered emitter arrays or strong coupling.

\begin{figure}
	\centering \includegraphics[width = \linewidth]{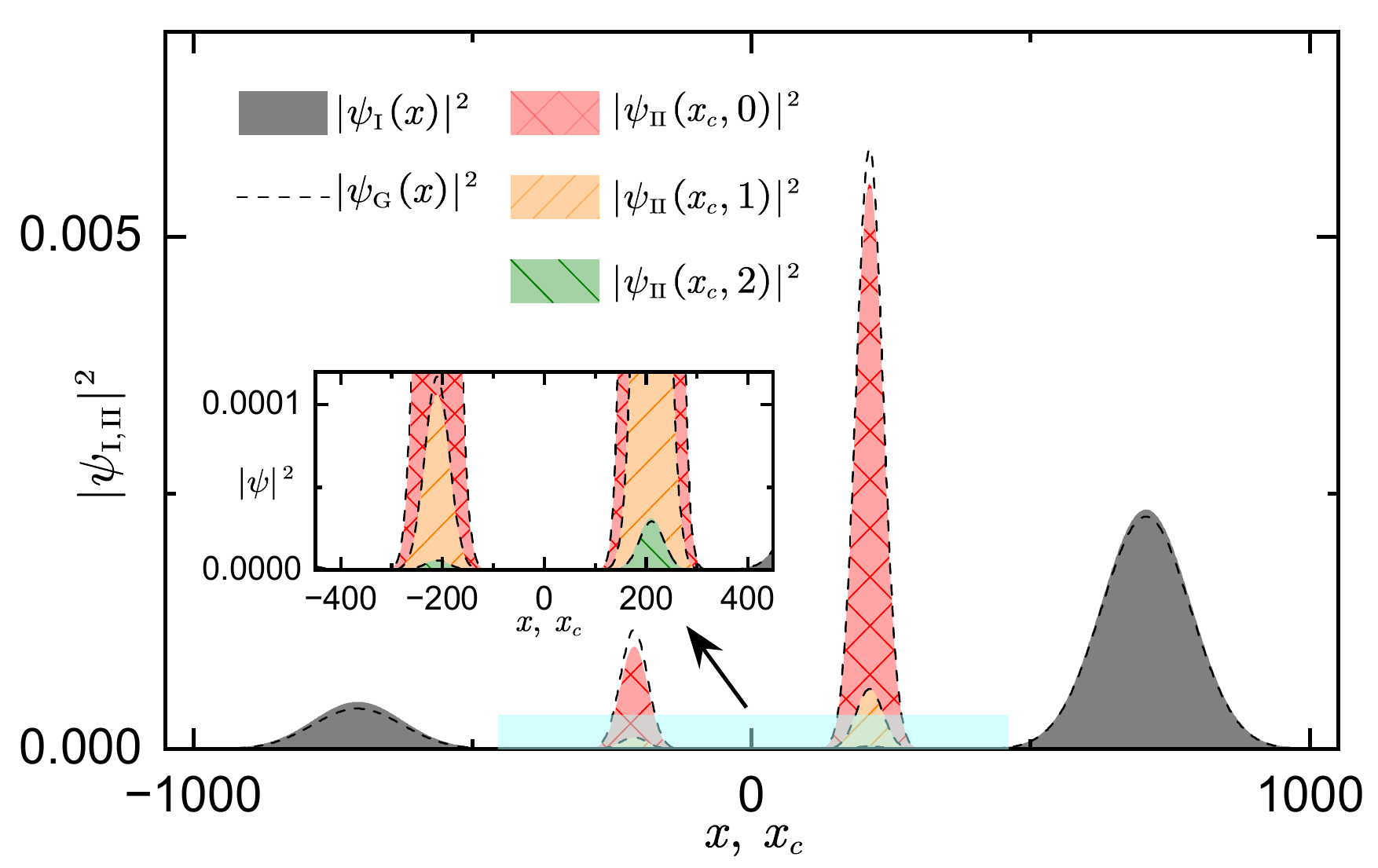}
	\caption{Spatial distribution of the single-photon field $|\psi_{\mathrm{I}}(x)|^2$ and the two-photon field $|\psi_{\mathrm{II}}(x_c,r)|^2$ at $Jt=600$ from Fig.~\ref{fig5}. Black solid curves represent analytical Gaussian wave packets [Eq.~(\ref{Gaussian_wave})], with spectral width $L_G$ for the single photons and $L_G v_{K_r} / v_{k_0}$ for the doublons. Amplitudes are $|t|^2$, $|r|^2$ and $|u_{\pm}|^2\times u(r)$ for the different components of the photon field.}
	\label{fig6}
\end{figure}

Moreover, the distinct group velocities shape the spatial profiles of the output states. The single-photon profile remains unchanged, while the doublon width is given by $W_D = v_K t_{\rm int} = W_S v_K / v_k$, where $W_S = v_k t_{\rm int}$ is the single-photon width and $t_{\rm int}$ is the interaction duration [see the inset of \figpanel{fig5}{b}]. Figure~\ref{fig6} displays the photonic field at $Jt = 600$, showing a clear spatial separation between the single-photon and two-photon components. The two-photon field exhibits strong spatial correlations, featuring a sharp peak at $r = 0$ and rapid decay~\cite{Winkler2006,Wang2020,Wang2024}. After the scattering, the single-photon and doublon output states inherit the spatial profile of the incident wave packet, centered at $x = v_{k_0/K_r} t$ with amplitudes given by $|t|^2$, $|r|^2$, and $|u_{\pm}|^2u_{K_r}(r)$, respectively.


\subsection{Modulating the scattering process with a real giant emitter}

\begin{figure}
	\centering \includegraphics[width = \linewidth]{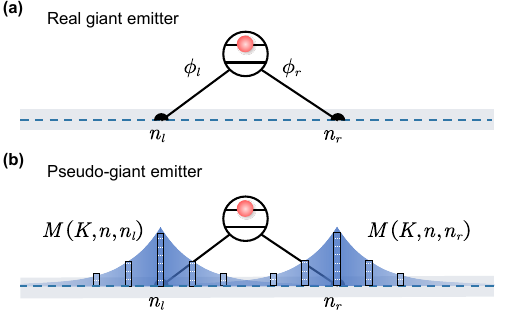}
	\caption{Layout with an RGA.
	(a) The emitter couples to the waveguide at two points $n_{l/r}$ with encoded local phase $\phi_{l/r}$, forming an RGA. 
	(b) In the PGA picture, sets of pseudo-coupling points form around the coupling positions $n_l$ and $n_r$.}
	\label{fig7}
\end{figure}

Real giant atoms, which couple to the waveguide at multiple points~\cite{FriskKockum2020, Gustafsson2014, Anton2014, Kockum2018, Kannan2020}, possibly with engineered local phases~\cite{Chen2022, Joshi2023}, provide enhanced control over light--matter interactions through interference among the coupling points. Leveraging this tunability, we propose an RGA configuration to achieve higher conversion efficiency into doublons and optimal unidirectional distribution. For simplicity and without loss of generality, we consider a giant emitter that couples to the waveguide at two points, $n_l$ and $n_r$, with local phases $\phi_l$ and $\phi_r$, as shown in \figpanel{fig7}{a}. The distance between the two coupling points is $d = |n_l - n_r|$. The initial interaction Hamiltonian is then expressed as
\begin{equation}
H_{\rm int} = g \sigma_- \mleft( e^{i \phi _l} a_{n_l}^\dag + e^{i \phi _r} a_{n_r}^\dag \mright) + \text{H.c.}
\end{equation}

Following the derivation in \secref{pseudo_giant_atom}, the effective interaction Hamiltonian becomes
\begin{align}
& H_{\rm int} = \sum_{K, n} \sum_{\tau = l, r} g e^{i \phi_\tau} M (K, n, n_\tau) D_K^\dag \sigma_-a_n + \text{H.c.} 
\notag \\
& = \sum_{K, n} \sum_{\tau = l, r} \frac{g_{|n - n_\tau|}}{\sqrt{N}} e^{- i K (n + n_\tau) / 2} e^{i \phi_\tau} D_K^\dag \sigma_- a_n + \text{H.c.} ,
\label{H_int_c_giant}
\end{align}
where $g_{|n - n_\tau|} = \sqrt{2} u_0 g e^{- |n - n_\tau| / L_u(K)}$. The physical process can be interpreted as follows: an incident photon at position $n$ combines with the photon emitted from coupling point $\tau$ at position $n_\tau$, incorporating the local phase $\phi_\tau$, to jointly excite a doublon in mode $K$. This excitation is characterized by an effective coupling strength $g_{|n - n_\tau|}$ and an accumulated propagation phase $\Phi_d = - K (n + n_\tau) / 2$, which depends on the center of mass of the two photons~\cite{Wang2024}. Note that Eq.~\eqref{H_int_c_giant} can be readily extended to the case of $N$ coupling points, where $\tau = 1, 2, \ldots, N$.

Comparing Eq.~\eqref{H_int_c_giant} with the small-atom case in Eq.~\eqref{interaction_H_D}, we note that, using the PGA picture, pseudo-coupling points now form around each coupling point, with additional local phases $\phi_{l/r}$, as shown in \figpanel{fig7}{b}. These pseudo-coupling points, associated with the coupling positions $n_l$ and $n_r$, introduce new interference pathways for doublon generation. When a single photon is located at $n$, the doublon can be generated via either the left ($l$) or right ($r$) coupling, corresponding to two distinct channels. In contrast to the case of a small emitter, where a single dominant channel prevails ($g_{n=0} \gg g_{n \neq 0}$), here both channels exhibit comparable strengths: $g_{|n_r - n_{\tau = r}|} = g_{|n_l - n_{\tau = l}|} = g$. Note that the coupling strength for each coupling point in Eq.~\eqref{H_int_c_giant} is initially uniform, but can be individually tuned by setting different values for $g_l$ and $g_r$, enabling enhanced modulation flexibility. In this scenario, it is possible to engineer the multipoint coupling to realize higher conversion efficiency and optimal interference condition to achieve $u_+ \gg u_- \simeq 0$, enabling the unidirectional generation of a doublon.

\begin{figure*}
	\centering \includegraphics[width=\linewidth]{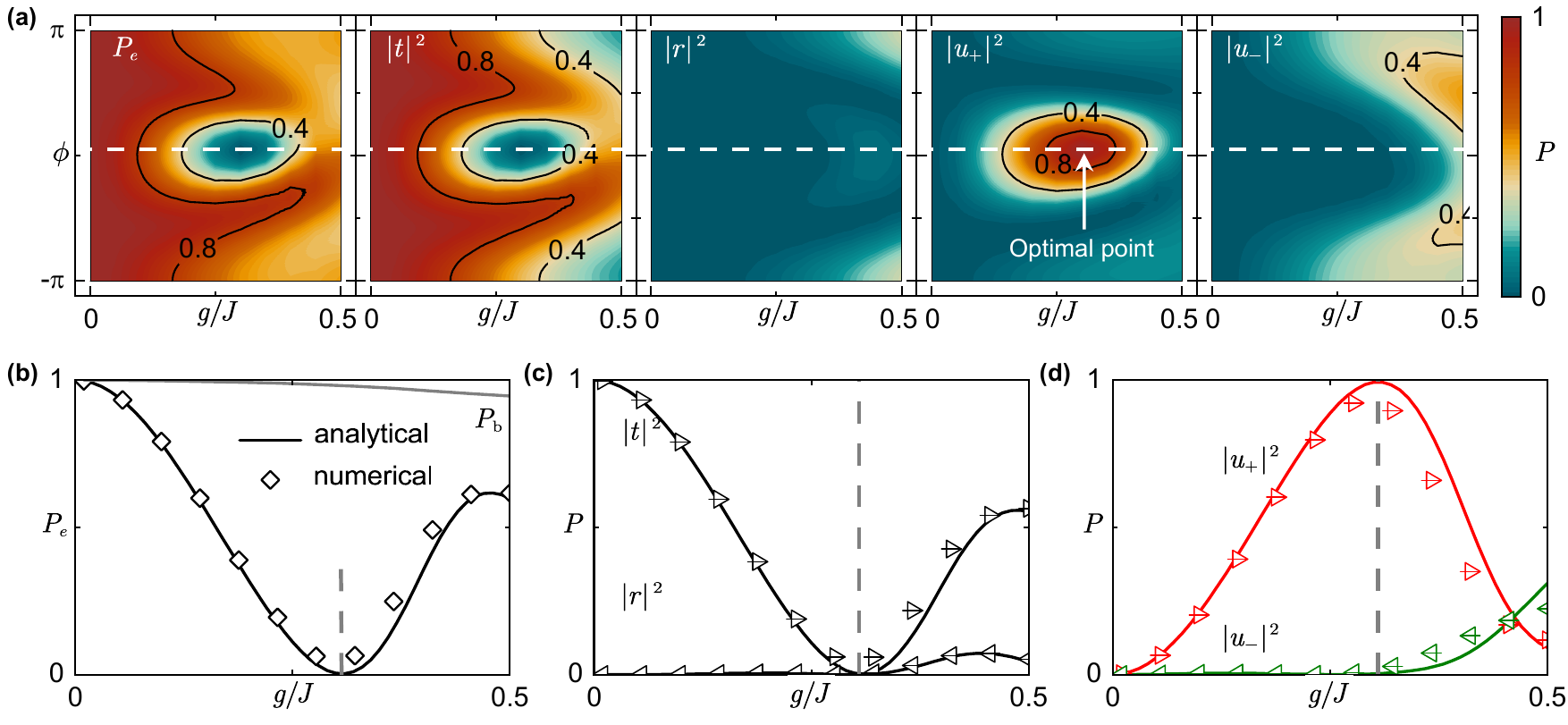}
	\caption{Finding optimal parameters for conversion to doublons.
	(a) The physical quantities $P_e$, $|t|^2$, $|r|^2$, and $|u_\pm|^2$ as functions of coupling strength $g$ and local phase $\phi$, for a giant emitter with coupling strengths $\{g, g, g\}$ and coupling phases $\{-\phi, 0, \phi\}$.
	(b--d) The lower panels show cross-section cuts along the white lines in the upper panels, comparing numerical simulations (diamond and triangular markers) with analytical solutions (solid curves). An optimal point exists at $g = 0.31J$ and $\phi = 0.05\pi$, where the incident single photon is fully converted into a right-going doublon. This point is indicated by the white arrow in the upper panels and dashed vertical gray lines in the lower panels. All other parameters are the same as in Fig.~\ref{fig5}.}
	\label{fig8}
\end{figure*}

By extending the derivation of the Lippmann--Schwinger equations to the giant-emitter interaction Hamiltonian in Eq.~\eqref{H_int_c_giant}, we identify an optimal interference regime in which the emitter couples to the waveguide at three points $\left\{-1,0,1\right\}$ with equal coupling strengths $\left\{g,g,g\right\}$ and encoded phases $\left\{-\phi,0,+\phi\right\}$. In this configuration, \figpanel{fig8}{a} shows the population of key physical quantities as functions of $g$ and $\phi$. At the parameter values indicated by the white arrow, the conversion efficiency approaches unity, signifying complete conversion of the incident single photon into a doublon state, accompanied by full emitter relaxation and vanishing single-photon transmission and reflection. \textit{Theoretically, the doublon field exhibits unidirectional propagation}, with $|u_{+K_r}|^2\simeq1$ and $|u_{-K_r}|^2\simeq0$, indicating optimal chirality in the output. Both the conversion efficiency and the chiral factor can be smoothly tuned across the full range $(0,1)$ by varying the coupling parameters $g$ and $\phi$.

\begin{figure}
	\centering \includegraphics[width = \linewidth]{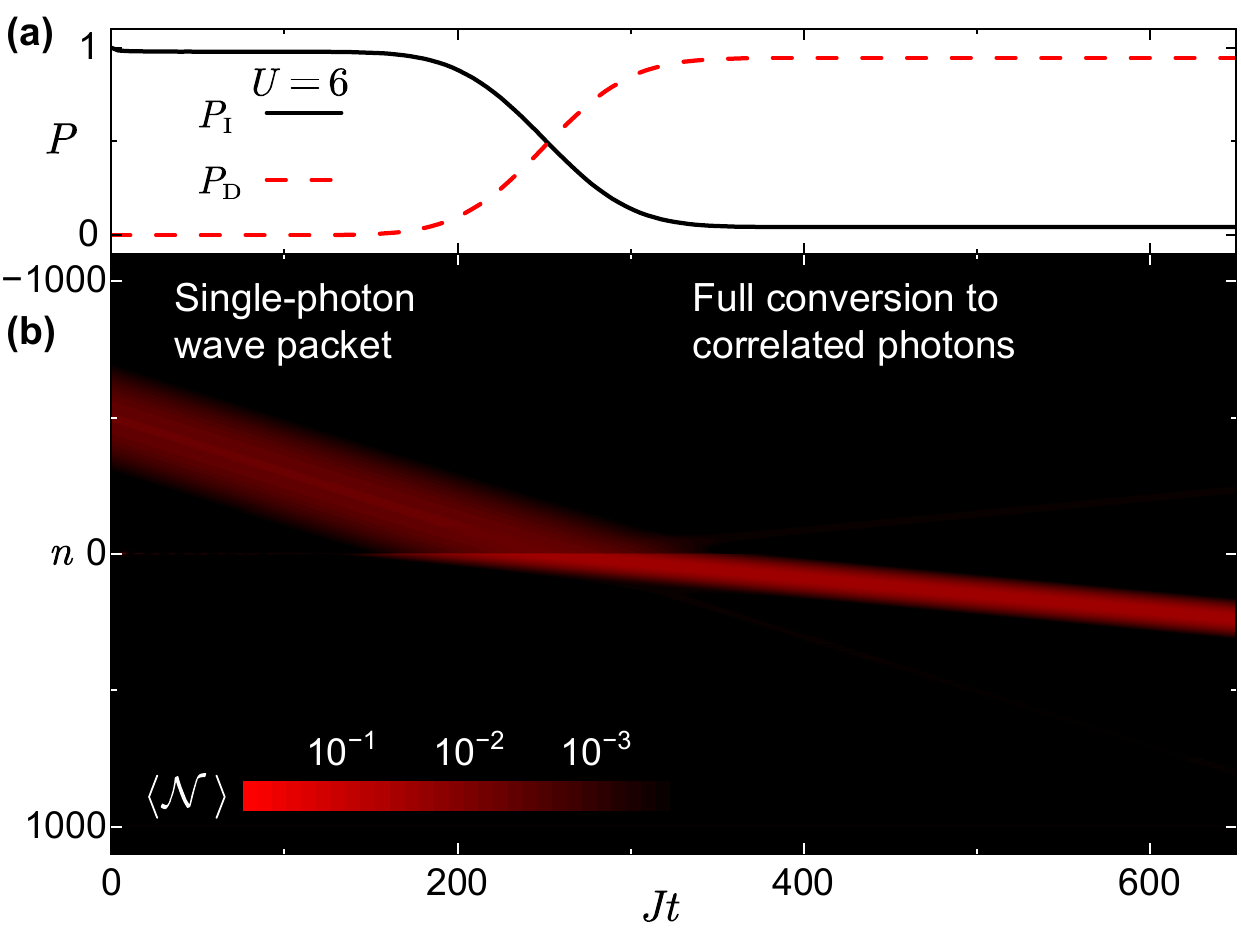}
	\caption{Conversion to a doublon.
	(a) Populations of the single- and two-photon states.
	(b) Photon number $\langle \mathcal{N} \rangle$ as a function of position $n$ and time $Jt$. Following scattering, the single photon is fully converted into a right-going doublon. Parameters correspond to the optimal operating point found in Fig.~\ref{fig8} ($g = 0.31J$, $\phi = 0.05\pi$), and are otherwise the same as in Fig.~\ref{fig5}.}
	\label{fig9}
\end{figure}

\Figurepanel{fig8}{b} compares numerical simulation results with the analytical solution, using the parameters indicated by the white dashed lines in \figpanel{fig8}{a}. The analytical predictions match the numerical data well. Using the optimal coupling parameters identified in Fig.~\ref{fig8}, we simulate the time evolution of the photonic field, as shown in Fig.~\ref{fig9}. The incident single photon is unidirectionally converted into a doublon, akin to optical refraction into a slow-light mode with a low group velocity. The simulated conversion efficiency reaches $94\%$, with negligible single-photon transmission and reflection. The mismatch with the theoretical results is caused by the initial bound state, as shown by the gray solid curve in \figpanel{fig8}{b}. In Appendix~\ref{Appendix_B}, when a Lorentzian wavepacket is incident instead, we achieve conversion efficiency identical to the Gaussian case. Moreover, by tuning the coupling parameters, the scattering dynamics can be fully controlled, enabling continuous modulation of the conversion efficiency and backscattering-free, cascaded correlated photon-pair generation.


\section{Cascaded scattering process: generating spatially separated correlated $N$-photon states}
\label{high_order}

In the Bose--Hubbard model, higher-order multi-photon bound states beyond the doublon are also present~\cite{Mansikkamaki2022}. As the photon number in these correlated states increase, the interaction-induced energy shift becomes larger, and the effective mass grows accordingly. By leveraging the mediation role of the emitter, \textit{the generated doublon can further scatter into a three-photon bound state, a triplon}, which in turn may scatter into even higher-order states, enabling a cascaded nonlinear scattering process. Moreover, each successive bound state exhibits a heavier effective mass and a slower group velocity. This velocity hierarchy leads to natural spatial separation among the different photon-number components after propagation. The resulting state is a spatiotemporally structured multiphoton entangled state, with distinct groups of photons propagating at different speeds and localized in different regions of space.


\subsection{Triplon generation}

In this section, we further investigate a doublon wave packet scattering off a second excited emitter, leading to the formation of a three-photon bound state, a triplon, as shown in \figpanel{fig1}{a}. The triplon band lies below the doublon bands, with a larger energy shift of $3U$ from the single-photon band. We consider the energy configuration illustrated in Fig.~\ref{fig1}, where the frequency of the second emitter $\Delta_{e2}$ satisfies the resonance condition $E_{\mathcal{K}_r} = \Delta_{e2} + E_{K_r}$,
with $E_{\mathcal{K}}$ denoting the triplon dispersion relation and $\mathcal{K}$ the triplon momentum. The second emitter couples to the waveguide at $n_2$ with coupling strength $g_2$. 

Following the derivation in the previous section, the interaction Hamiltonian for the triplon generation can be written as
\begin{equation}
H_{\rm int} = g_2 \sum_{\mathcal{K}, K} M^* (\mathcal{K}, K, n_2) T_\mathcal{K}^\dag D_K \sigma_- ,
\end{equation}
where $T^\dag_\mathcal{K}$ is the creation operator for the triplon state $\mathcal{K}$. The transition-matrix element is $M (\mathcal{K}, K, n_2) = \langle \text{vac}|D_K a_{n_2} T_\mathcal{K}^\dag |\text{vac}\rangle$, between the incoming state $\ket{K, n_2}$ and the outgoing triplon state $\ket{\mathcal{K}}$. The transition rate is proportional to the overlap between two three-photon states: one is the correlated three-photon state; the other consists of a doublon state $K$ and a single photon located at $n_2$ emitted by the second emitter. During the scattering process where the doublon is converted into a triplon, the localized coupling of emitter $2$ behaves as a PGA, due to the spatial extent of the triplon wave function.

\begin{figure*}
	\centering \includegraphics[width=\linewidth]{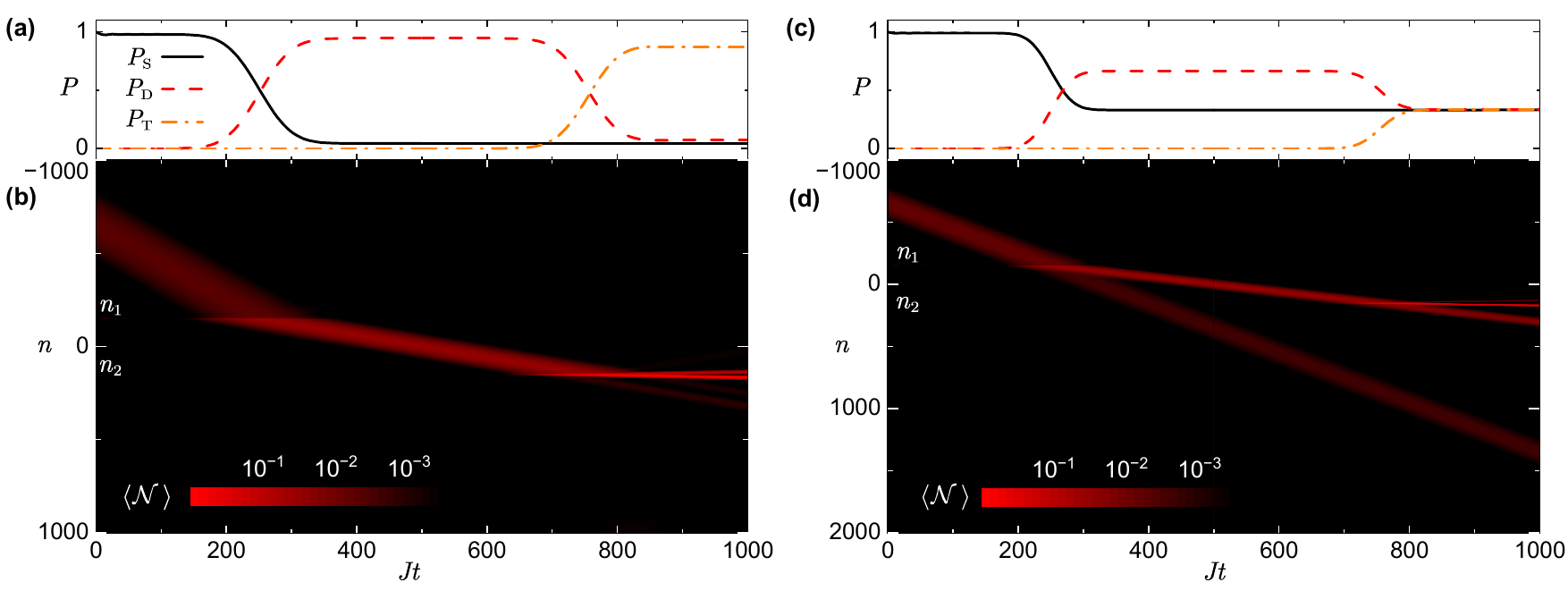}
	\caption{Cascaded scattering process in a nonlinear waveguide.
	(a, c) Populations of the single-photon, doublon, and triplon states.
	(b, d) Photon number $\langle \mathcal{N}\rangle$ as a function of position $x$ and time $Jt$. 
	The frequency of the second emitter is $\Delta_{e2} = -11.869J$. Emitter positions are fixed at $n_1 = -150$ and $n_2 = 150$. Parameters for (a, b): $g_1 = 0.31J$, $\phi_1 = 0.05 \pi$, $g_2 = 0.042J$, $\phi_2 = 0.167 \pi$, and $L_G = 0.002$. Parameters for (c,d): $g_1 = 0.202J$, $\phi_1 = 0.05 \pi$, $g_2 = 0.0255J$, $\phi_2 = 0$ and $L_G = 0.003$. All other parameters are the same as in Fig.~\ref{fig9}.}
	\label{fig10}
\end{figure*}

In Fig.~\ref{fig10}, we present numerical simulations of this second scattering stage, building upon the initial state generated in the first stage (see Fig.~\ref{fig9}). To enhance the triplon conversion efficiency, the second emitter is implemented as an RGA, coupling to the waveguide at three points $\{n_2 - 1, n_2, n_2 + 1\}$ with equal coupling strength and phase configuration $\{\phi, 0, \phi\}$. By jointly optimizing the coupling parameters of both giant emitters, we achieve a nearly complete cascade: an incident single photon ($S$) is first converted into a two-photon bound state (doublon $D$), which is subsequently converted into a three-photon bound state (triplon $T$), as shown in \figpanel{fig10}{a} and \figpanel{fig10}{b}. This stepwise energy transfer across multiple quantum nonlinear stages resembles optical refraction across a series of media, with each mediating a mode conversion.

Moreover, the conversion efficiency for doublons and triplons can be continuously tuned by adjusting the coupling parameters of the giant-atom setup. In \figpanel{fig10}{c} and \figpanel{fig10}{d}, we observe an approximately equal partition of the photon population among the single-, two-, and three-photon subspaces. As the photon number increases, the effective mass of the bound state grows and its group velocity decreases accordingly. At late times, the three wave-packet components (the single-photon, doublon, and triplon states) are \textit{naturally separated in space}. 

Note that the energy $\omega_{k_0} + \Delta_{e2}$ does not resonate with any eigenmode of the nonlinear waveguide. Consequently, the second emitter cannot couple the incident single photon to any photonic excitation. Although part of the single photon transmits through the first emitter, it propagates freely past the second emitter without significant scattering. In contrast, the second emitter and the incoming doublon state satisfy the resonance condition $E_{\mathcal{K}_r} = \Delta_{e2} + E_{K_r}$. This selective interaction mechanism enables the emitter to act as a ``scattering trigger'', activated only upon arrival of a correlated two-photon state. 

By harnessing multi-photon bound states in the nonlinear waveguide and the engineered coupling of RGAs, the system realizes a deterministic cascaded scattering process, $S \to D \to T$. Given the existence of higher-order bound states in such systems, this sequential nonlinear conversion can, in principle, be extended to generate increasingly complex correlated photonic states through sequential nonlinear conversion. This architecture enables the deterministic encoding of quantum information into propagating multiphoton states, ``flying qubits'', with enhanced correlation depth, paving the way toward scalable generation of entangled photonic clusters.


\subsection{Spatial and temporal entanglement}

\begin{figure}
	\centering \includegraphics[width = \linewidth]{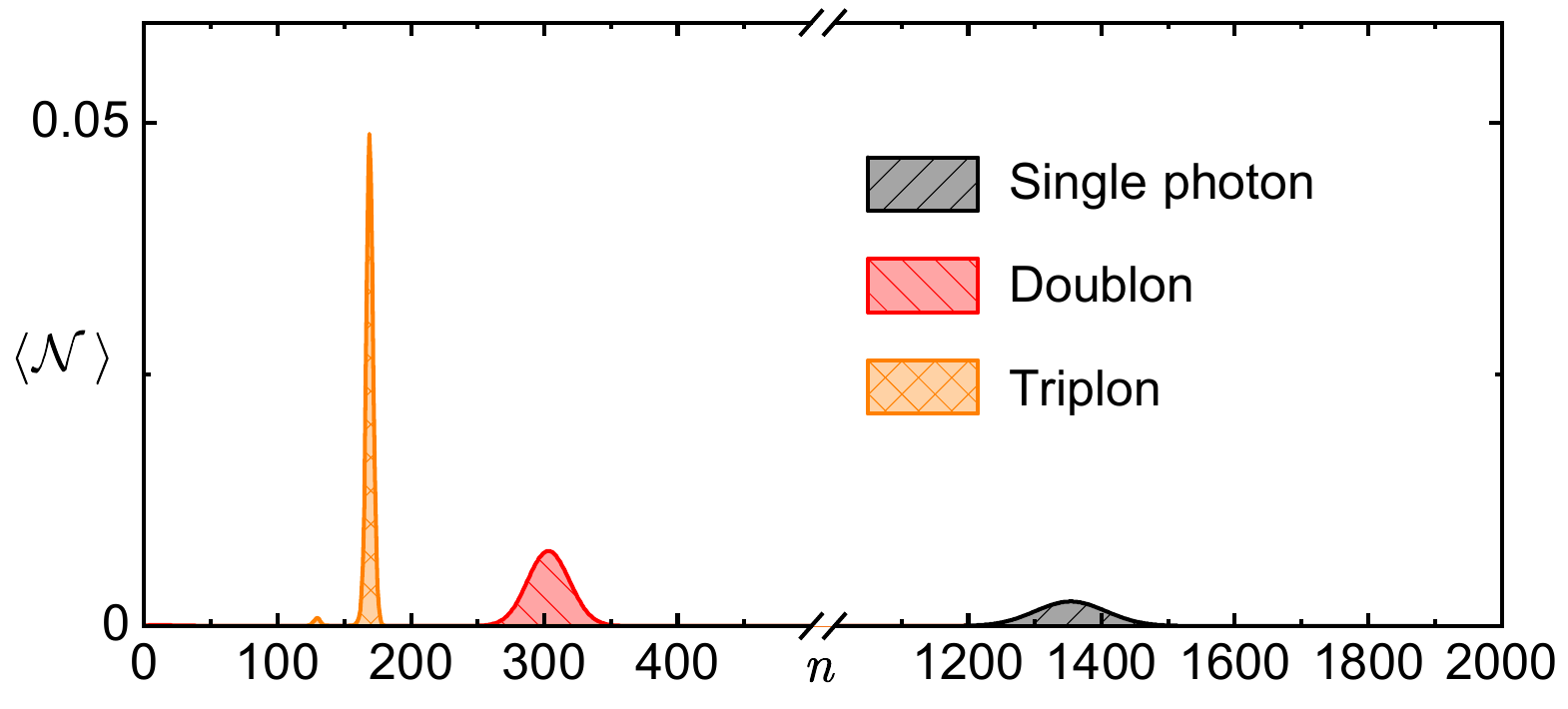}
	\caption{Output spatiotemporal entanglement of the photonic field at $Jt=1000$ from \figpanel{fig10}{d}. The single-photon, doublon, and triplon wave packets are spatially separated due to their photon-number-dependent group velocities, forming a multipartite quantum state with entanglement between arrival time and photon number.
	}
	\label{fig11}
\end{figure}

The cascaded scattering process culminates in a spatiotemporally ordered quantum superposition, where the incident single photon, doublon, and triplon automatically separate into distinct space-time bins due to their photon-number-dependent group velocities. As derived analytically and visualized in Fig.~\ref{fig11}, the final state evolves into
\begin{equation}
\ket{\Psi_{\rm out}} \simeq \alpha \ket{S}_{\tau_1} + \beta \ket{D}_{\tau _2} + \gamma \ket{T}_{\tau _3} ,
\end{equation}
where 
\begin{align}
	\ket{S} &\equiv a_k^\dag \ket{\text{vac}, e, e} , 
	\\
	\ket{D} &\equiv D_K^\dag \ket{\text{vac}, g, e} ,
	\\
	\ket{T} &\equiv T_\mathcal{K}^\dag \ket{\text{vac}, g, g} ,
\end{align}
represent the single-photon, doublon, and triplon wavepacket states, respectively. Here, the temporal ordering $\tau_1 < \tau_2 < \tau_3$ arises from the hierarchy of group velocities $v_g^{(1)} > v_g^{(2)} > v_g^{(3)}$, where the superscripts denote the photon number in each bound state. The spatial distribution of the doublon and triplon, characterized by center-of-mass coordinates $x_c$ and $X_c$, allow their interpretation as composite bosonic quasiparticles. Additionally, the giant-atom coupling architecture enables flexible control of the superposition amplitudes $\alpha$, $\beta$, and $\gamma$, allowing programmable control of the output state.

This automatic spatiotemporal sorting transcends conventional time-bin entanglement schemes, which typically rely on time-dependent modulation pulses, multilevel emitters~\cite{Gheri1998, Schon2005, Pichler2017, Kannan2020_2, Besse2020, Wein2022, Tiurev2022}, or structured multimode baths~\cite{Vega2023}. In contrast, our architecture harnesses strongly correlated multi-photon bound states as intrinsic carriers of quantum information, enabling ``correlated multi-photon flying qubits" with enhanced encoding capacity. The intrinsic non-classicality of the output state manifests through three distinct yet interrelated features: 
\begin{enumerate}
\item \textbf{Photon-number superposition across temporal modes:} The state exhibits a coherent superposition of different photon numbers in distinct space-time bins, reflecting the temporal delocalization of quantum excitations.
\item \textbf{Multi-photon bunching within bound-state wavepackets:} Each localized wavepacket contains a well-defined multiphoton bound state, exhibiting strong internal photon-photon correlations, evidenced by bunching in the relative coordinates, within a single temporal mode.
\item \textbf{Entanglement between arrival time and photon number:} The final state is non-separable in the temporal and photonic number degrees of freedom, realizing entanglement between \textit{when} the excitation arrives and \textit{how many} photons it contains.
\end{enumerate}
Harnessing the innate correlations in the spatial, temporal, and photon-number  dimensions of the output states, this hybrid encoding scheme can multiplex them into a powerful framework for high-dimensional quantum information processing.


\section{conclusion}
\label{sectionVI}

We have established a paradigm for generating propagating multi-photon entangled states through cascaded inelastic scattering in a nonlinear waveguide. The central point of this work is the discovery that a far detuned TLE can mediate photon-number upconversion ($S\to D\to T$), bypassing the limitation of resonant scattering. This process is governed by an emergent non-local scattering potential, theoretically captured by the pseudo-giant atom formalism that we introduced, which reveals how the nonlinear multi-photon state reshapes the light--matter interaction.

By implementing this architecture with real giant atoms, we demonstrated unidirectional, high-efficiency conversion and programmable state synthesis, culminating in a self-sorted superposition of spatially and temporally isolated photon-number states. The resulting spatiotemporally entangled multi-photon state opens immediate pathways toward quantum-enhanced metrology through deterministic NOON-state synthesis~\cite{Compagno2017, Dengis2025}, distributed quantum computing using time-bin-encoded multi-photon graph states~\cite{Penas2024}, and quantum state engineering via natural photon-number redistribution. More broadly, it bridges the gap between quantum nonlinear optics and correlated many-body physics, offering a scalable platform to explore exotic photonic matter.

This proposal is readily implementable in state-of-the-art superconducting 
circuits. An array of capacitively coupled transmons, whose intrinsic 
anharmonicity naturally provides the required nonlinear 
potential~\cite{Blais2021}, can serve as the nonlinear waveguide considered 
in this work~\cite{Karamlou2024,Claudia2025}. Giant-atom architectures in 
superconducting circuits featuring multiple connection points, have already 
been experimentally demonstrated~\cite{Kannan2020,Joshi2023}. Our results 
encourage further investigation of similar phenomena in nonlinear continuous 
waveguides and pave the way for future studies of nonlinear dynamics in 
various systems with photon-photon interactions. 


\begin{acknowledgments}
We are grateful to Dr. Tao Shi for useful discussions about scattering 
theory. The quantum dynamical simulations were performed using the 
open-source tools QuTiP~\cite{Johansson12qutip, Johansson13qutip, 
Lambert2024} and QuSpin~\cite{Weinberg2017, Weinberg2019}. X.W.~is supported 
by the National Natural Science Foundation of China (NSFC) (Grant 
No.~12174303). A.F.K. acknowledges support from the Swedish Foundation for 
Strategic Research (grant numbers FFL21-0279 and FUS21-0063), the Horizon 
Europe programme HORIZON-CL4-2022-QUANTUM-01-SGA via the project 101113946 
OpenSuperQPlus100, and from the Knut and Alice Wallenberg Foundation through 
the Wallenberg Centre for Quantum Technology (WACQT).

\end{acknowledgments}


\appendix

\section{Derivations in momentum space}
\label{momentum_space_derivation}

In this section, we derive the scattering state in momentum space, which is formally equivalent to the real-space treatment presented in \secref{real_space_scattering}. Applying a Fourier transform to convert the photon position coordinate $x$ into the wave vector $k$, the interaction Hamiltonian is
\begin{align}
& H_{\rm int} = \sum_K \sum_k g M^* (K, k, 0) D_K^\dag \sigma_- a_k + \text{H.c.}, 
\label{H_int_momentum}
\\
& M (K, k, 0) = \frac{\sqrt{2}}{N} \sum_m e^{- i k m} e^{i K m / 2} u_0 e^{- |m| / L_u(K)},
\label{M_transition_k}
\end{align}
and the scattering-state solution has the form
\begin{equation}
\ket{\Psi (t)} = \mleft[ \sum_k c_{e, k} (t) \sigma_+ a_k^\dag + \sum_K c_K (t) D_K^\dag \mright] \ket{\mathrm{vac}} , 
\label{scattering_state_momentum}
\end{equation}
where $c_{e, k}(t)$ denotes both the emitter and the single-photon state $k$ being excited. For notational simplicity, we suppress the emitter label, so that $c_{e,k} \rightarrow c_{k}$. 

From the secular equation, we obtain the following system of equations:
\begin{align}
& \hspace{-0.2cm} \mleft( \omega _k + \Delta _e \mright) c_k + g \sum_K M (K, k, n_0) c_K = \mleft( \omega_{k_0} + \Delta_e \mright) c_k ,
\\
& E_K c_K + g \sum_k M^* (K, k, n_0) c_k = \mleft( \omega_{k_0} + \Delta_e \mright) c_K .
\end{align}
Note that these equations describe \textit{the coupling between two momentum modes $k$ and $K$}. They can be rearranged into the form
\begin{align}
c_k &= \delta_{k, k_0} + g \sum_K \frac{M (K, k, n_0) c_K}{\omega_{k_0} - \omega_k + i 0^+} ,
\label{c_k_initial_state}
\\
c_K &= \frac{g \sum_k M^* (K, k, n_0) c_k}{E_{K_r} - E_K + i 0^+} ,
\label{c_K_initial_state}
\end{align}
where $\delta_{k, k_0}$ indicates that the initial state is a right-going single photon. 

By defining~\cite{Shi2018}
\begin{equation}
C_0(m) = \sum_k e^{i k m} c_k ,
\label{eq:def_C0}
\end{equation}
we obtain a complex convolution equation
\begin{align}
& C_0 (m_0) = e^{i k_0 m_0} - \frac{1}{v_{k_0} v_{K_r}} \times
\notag \\
&  \sum_{m_1, m_2} g_{m_1} g_{m_2} e^{i K_r |m_1 - m_2| / 2} e^{i k_0 |m_0 - m_1|} C_0 (m_2) , 
\label{C_0_equations}
\end{align}
where 
\begin{equation}
g_m = \sqrt{2} u_0 g e^{- |m| / L_u(K_r)}
\end{equation}
denotes the effective coupling strength per pseudo-coupling point of the PGA. The first term in Eq.~\eqref{C_0_equations}, $\exp(i k_0 m_0)$, corresponds to the free propagation of the incident wave, while the second term captures the scatting disturbance due to the nonlocal potential. The convolution arises naturally from this nonlocality: the output $C_0(m_0)$ depends not only on the input at position $m_0$, but also on the responses $C_0(m_2)$ at neighboring sites. This formulation describes multiple scattering events and reflects interference among different scattering channels.

Following the approach in \secref{real_space_scattering}, we find that the final state takes the form
\begin{align}
& \ket{f_i} = \ket{\Psi (t \rightarrow \infty)} \simeq
\notag \\
& \mleft[ \sum_{\pm k} c_{e, \pm k} (+ \infty) \sigma_+ a_{\pm k}^\dag + \sum_{\pm K} u_{\pm K} (+ \infty) D_{\pm K}^\dag \mright] \ket{\mathrm{vac}},
\end{align}
where the momentum components are explicitly separated into positive and negative branches to distinguish the right- and left-going single-photon waves and the generation of doublon waves. After obtaining the scattering state for a single-photon plane-wave input, we now return to Eq.~(\ref{LS_equation_f}) to derive the output state by projecting the final state onto the subspaces spanned by $\ket{e, k}$ and $\ket{K}$, respectively.


\subsection{Doublon component} 

We substitute the scattering state from Eq.~\eqref{scattering_state_momentum} and the interaction Hamiltonian from Eq.~\eqref{H_int_momentum} into Eq.~\eqref{LS_equation_f}, and project onto the right-going doublon state $\ket{+K}$, yielding
\begin{align}
\hspace{-0.1cm} \braket{+K | f} &= \braket{+K | \Psi_{\rm sc}} - \bra{+K} \frac{1}{E_{K_r} - H_0 - i 0^+} H_{\rm int} \ket{\Psi_{\rm sc}} 
\notag \\
&= \sum_{K'} \bra{+K} \frac{g \sum_k M^* (K', k, n_0) c_k}{E_{K_r} - E_{K'} + i 0^+} \ket{K'}
\notag \\
&- \sum_{K'} \bra{+K} \frac{g \sum_k M^* (K', k, n_0) c_k}{E_{K_r} - E_{K'} - i0^+} \ket{K'} . \label{f_project_onto_p_K_momentum}
\end{align}
Using the expression for $M$ in Eq.~\eqref{M_transition_k}, the definition of $C_0(m)$ in Eq.~\eqref{eq:def_C0}, and evaluating the integrals via the retarded and advanced Green's functions Eqs.~\eqref{Green_retarded} and \eqref{Green_advanced}, we simplify Eq.~\eqref{f_project_onto_p_K_momentum} to
\begin{align}
& \braket{+K | f} = \braket{+K_r | f} 
\notag \\
&= - i \frac{1}{v_{K_r}} \sum_m g_m e^{- i K_r m / 2} \mleft[ \Theta (-m) + \Theta (+m) \mright] C_0 (m) 
\notag \\
&= - i \frac{1}{v_{K_r}} \sum_m g_m e^{- i K_r m / 2} C_0 (m) .
\end{align}
Here, the retarded and advanced Green's functions, corresponding to the first and second terms in Eq.~\eqref{f_project_onto_p_K_momentum}, contribute $\Theta(-m)$ and $\Theta(+m)$, respectively. This temporal and spatial structure reflects the full scattering history: the initial state evolves into the scattering state under the retarded Green's function from $t = - \infty$ to $t = 0$, covering the negative spatial part of the nonlocal potential. The final state evolves back under the advanced Green's function from $t = + \infty$ to $t = 0$, accounting for the positive spatial part. Together, these contributions encode the complete scattering process across both time ($t = - \infty \to t = 0 \to t = + \infty$) and space ($x = - \infty \to x = 0 \to x = + \infty$), as captured by the combination $\Theta(-x) + \Theta(+x)$. Thus, the doublon components of the final state are
\begin{align}
&u_{+K_r} = \braket{+K_r | f} = - i \frac{1}{v_{K_r}} \sum_m g_m e^{- i K_r m / 2} C_0 (m) , 
\label{k_space_f_pK}
\\
&u_{-K_r} = \braket{-K_r | f} = - i \frac{1}{v_{K_r}} \sum_m g_m e^{+ i K_r m / 2} C_0 (m) . 
\label{k_space_f_mK}
\end{align}
%


\subsection{Single-photon component}

We now project the final state onto the right-going single-photon state $\ket{+k, e}$ (suppressing the emitter label, we denote this state as $\ket{k}$):
\begin{align}
\hspace{-0.05cm} \braket{+k | f} =& \braket{+k | \Psi_{\rm sc}} - \bra{+k} \frac{1}{E - H_0 - i 0^+} H_{\rm int} \ket{\Psi_{\rm sc}} 
\notag \\
=& \braket{+k | +k_0} + \sum_k \bra{+k} \frac{g \sum_K M (K, k, n_0) c_K}{\omega_{k_0} - \omega_k + i 0^+} \ket{k}
\notag \\
& - \sum_k \bra{+k} \frac{g \sum_K M(K, k, n_0) c_K}{\omega_{k_0} - \omega_k - i 0^+} \ket{k} .
\label{f_project_k_1}
\end{align}
Here, the first term corresponds to the initial state [i.e., the $\delta_{k,k_0}$ in Eq.~(\ref{c_k_initial_state})]. 

The Green's function integrals over $k$ are
\begin{align}
& \frac{1}{N} \sum_k \frac{e^{- i k m / 2}}{\omega_{k_0} - \omega_k + i 0^+} \ket{k} = - \frac{i}{v_k} \times
\notag \\
& \mleft[ e^{- i k_0 m / 2} \Theta (-m) \ket{+k_0} + e^{i k_0 m / 2} \Theta (+m) \ket{-k_0} \mright],
\\
& \frac{1}{N} \sum_k \frac{e^{- i k m / 2}}{\omega_{k_0} - \omega_k - i 0^+} \ket{k} = + i \frac{1}{v_k} \times
\notag \\
& \mleft[ e^{- i k_0 m / 2} \Theta (+m) \ket{+k_0} + e^{i k_0 m / 2} \Theta (-m) \ket{-k_0} \mright].
\end{align}
Thus, Eq.~(\ref{f_project_k_1}) is
\begin{equation}
\braket{+k_0 | f} = 1 + \sum_{m_1} \frac{-i}{v_k} g_{m_1} e^{- i k_0 m_1} \sum_K e^{i K m_1 / 2} c_K .
\end{equation}
The summation over $K$ can be evaluated using Eq.~\eqref{c_K_initial_state}:
\begin{equation}
\sum_K e^{i K m_1 / 2} c_K = \sum_{m_2} \frac{-i}{v_{K_r}} e^{i K_r |m_1 - m_2| / 2} g_{m_2} C_0 (m_2) .
\end{equation}
The final state amplitudes for the single-photon components are then
\begin{align}
& t = \braket{+k_0, e | f} = 1 -
\notag \\
& \frac{1}{v_{k_0} v_{K_r}} \sum_{m_1, m_2} g_{m_1} g_{m_2} e^{- i k_0 m_1} e^{i K_r |m_1 - m_2| / 2} C_0 (m_2) , 
\label{k_space_f_t}
\\
& r = \braket{-k_0, e | f} = 
\notag \\
& - \frac{1}{v_{k_0} v_{K_r}} \sum_{m_1, m_2} g_{m_1} g_{m_2} e^{i k_0 m_1} e^{i K_r |m_1 - m_2| / 2} C_0 (m_2) . 
\label{k_space_f_r}
\end{align}

Ultimately, the final outgoing state, given an initial incident single-photon plane wave $\exp(i k_0 x)$, is
\begin{align}
\ket{i} &= \ket{e, +k_0} ,
\\
\ket{f_i} &= t \ket{e, +k_0} + r \ket{e, -k_0} + u_{\pm K_r} \ket{\pm K_r} ,
\label{f_state}
\end{align}
where $t$ and $r$ denote the transmission and reflection amplitudes, respectively, of the single photon, and $u_{\pm K_r}$ is the probability amplitudes for the generated right- and left-going doublon modes. Note that, although the incident state only consists of a single-photon plane wave, the scattered state contains not only the reflection $r$ and transmission $t$ single-photon components, but also the right- and left-going doublon waves $\ket{\pm K_r}$. The final results can be obtained numerically by solving the complex convolution equation [Eq.~\eqref{C_0_equations}] for $C_0(m)$, and then substituting into Eqs.~\eqref{k_space_f_pK}, \eqref{k_space_f_mK}, \eqref{k_space_f_t}, and \eqref{k_space_f_r}.

\section{The wave-packet shape}
\label{Appendix_B}

In our analytical derivation, we assume the incident wave packet to be a plane wave, i.e., $\ket{\psi(t = 0)} = \exp(i k_0 x)$. For a specific single mode $k$, this assumption allows the computation of the corresponding scattering amplitudes, such as $t(k)$, $r(k)$, and $u_{\pm K_r}(k)]$, each of which depends on $k$. Similar to the Wigner--Weisskopf approximation, if the momentum spread of the wave packet is much smaller than the single-photon bandwidth ($L_G \ll 4J$), the dispersion near $k_0$ can be approximated as linear. In this regime, the narrow momentum distribution can be effectively treated as centered around $k_0$. Consequently, the scattering amplitudes can be approximated as constants: $[t(k_0)$, $r(k_0), u_{\pm K_r}(k_0)]$. This approximation is validated numerically in Fig.~\ref{fig3}.

\begin{figure}
	\centering \includegraphics[width = \linewidth]{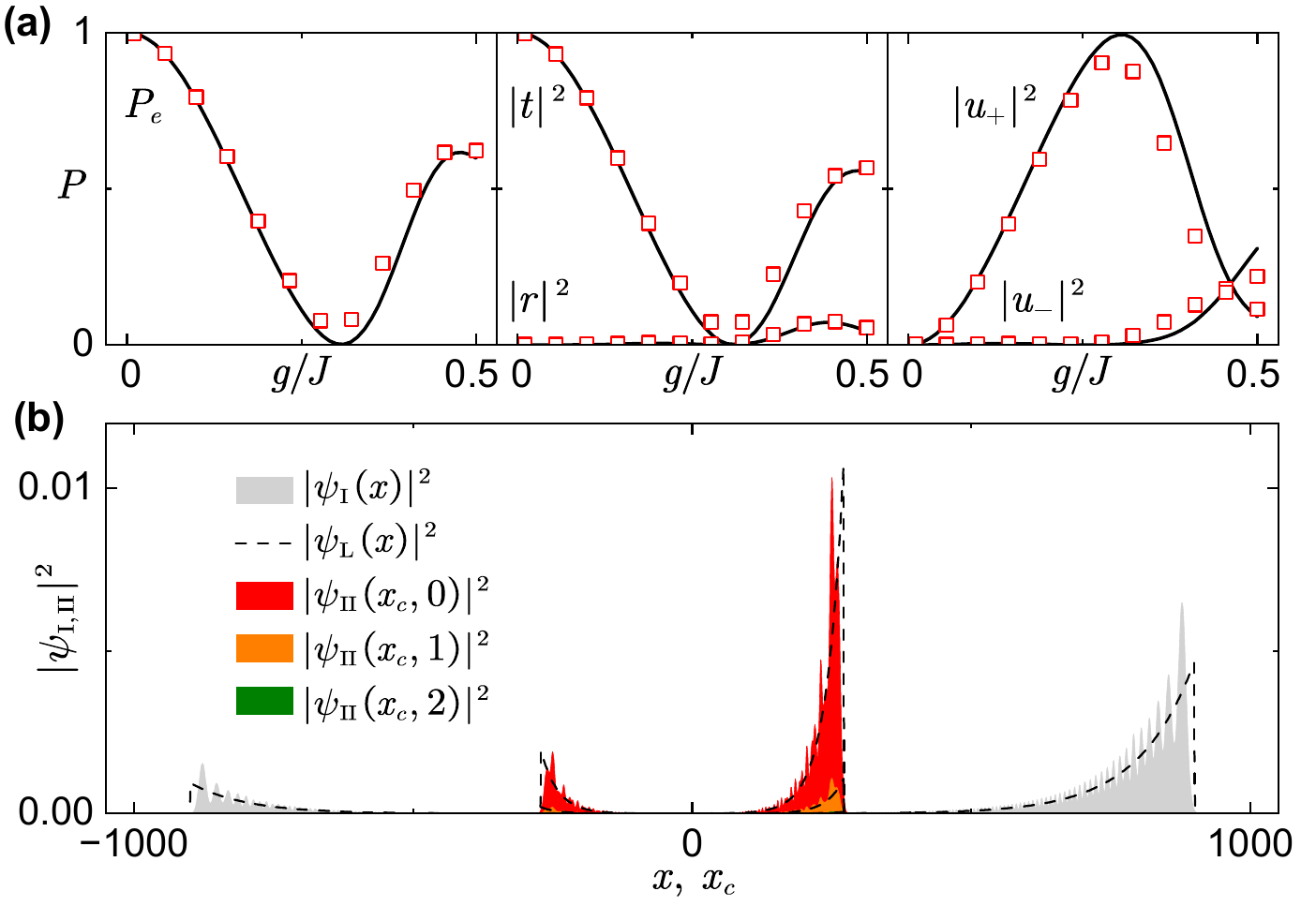}
	\caption{Scattering results for a Lorentzian wave packet.
	(a) The physical quantities as a function of $g$ with the initial Lorentzian wave packet shape from Eq.~\eqref{L_shape}.
	(b) The scattering field. The parameters are the same as in Fig.~\ref{fig8} in the main text.}
	\label{fig_s}
\end{figure}

We also perform numerical simulations using a Lorentzian wave packet. The initial state in momentum and real space is then given by
\begin{align}
\psi_L (k) &= \sqrt{\frac{L_L}{\pi}} \frac{1}{L_L - i (k - k_0)} , 
\notag \\
\psi_L (x) &= \sqrt{2 L_L} e^{i k_0 x} e^{x L_L} \Theta (-x_0) .
\label{L_shape}
\end{align}
Here, $L_{\mathrm{L}}$ denotes the spatial spread of the Lorentzian wave packet. Using the same simulation parameters as for the Gaussian wave packet discussed in the main text, we present the results in Fig.~\ref{fig_s}. After scattering, the generated doublon wave retains the profile of the incident photon wave packet, albeit with a modified bandwidth $W_D = (v_K / v_k) W_S$.



%

\end{document}